\documentclass[12pt,english]{article}
\usepackage{lmodern}
\usepackage[T1]{fontenc}
\usepackage[latin9]{inputenc}
\usepackage{amsmath}
\usepackage{amssymb}
\usepackage{esint}

\usepackage{geometry}
\makeatletter
\usepackage{float} 
\usepackage[colorlinks=true, linkcolor=ForestGreen, citecolor=blue]{hyperref}
\usepackage{cite}
\usepackage{babel}
\usepackage{mathdots}
\usepackage{amsfonts}
\usepackage{mathrsfs}
\usepackage{amsmath}
\usepackage{esint}
\usepackage{graphicx}
\usepackage{youngtab}
\usepackage{multicol}
\usepackage{multirow}
\usepackage{slashed}
\usepackage{simplewick}
\usepackage{braket}
\usepackage{bm}
\usepackage{relsize}
\usepackage[dvipsnames,table,xcdraw]{xcolor}
\usepackage{subfigure}
\usepackage{feynmp}
\usepackage{pifont}
\usepackage{cancel}
\numberwithin{equation}{section}
\allowdisplaybreaks

\makeatother

\usepackage[dvipsnames]{xcolor}

\begin{document}
\author{Wan-Zhe Feng\footnote{Email: vicf@tju.edu.cn},~~Zong-Huan Ye\footnote{Email: y2953083702@tju.edu.cn},~~Zi-Hui Zhang\footnote{Email: zhangzh\_@tju.edu.cn}\\
\textit{\small{Center for Joint Quantum Studies and Department of Physics,}}\\
\textit{\small{School of Science, Tianjin University, Tianjin 300350, PR. China}}}
\title{Supercool with PPO: Exploring Supercooled Phase Transitions via Reinforcement Learning}

\maketitle

\begin{abstract}
Gravitational waves from cosmological first-order phase transitions provide a powerful probe of hidden sectors and beyond the Standard Model physics. However, identifying phenomenologically relevant benchmark points remains computationally challenging, since viable and detectable signals typically occupy only a small fraction of the scanned parameter space. In this work, we introduce a reinforcement learning strategy based on Proximal Policy Optimization (PPO) to accelerate the search for gravitational wave signals from supercooled phase transitions in a minimal dark $U(1)_x$ sector. We construct a numerical reinforcement learning environment that maps the microscopic model parameters to the corresponding phase transition and gravitational wave observables, using a gauge-independent low-temperature formulation of the effective action. Several reward designs are developed to guide the agent toward parameter regions producing large gravitational wave amplitudes, broad frequency coverage, and detector sensitive benchmark points. We compare the PPO scans with conventional Monte Carlo scans in both narrow and broad windows of the $U(1)_x$ vacuum expectation value. Our results demonstrate that PPO provides an efficient goal-directed search strategy for gravitational wave phenomenology and offers a broadly applicable framework for learning-assisted exploration of high-dimensional scientific parameter spaces.
\end{abstract}

\newpage{}

{\hrule height 0.4mm \hypersetup{colorlinks=black,linktocpage=true}
\tableofcontents
\vspace{0.5cm}
\hrule height 0.4mm}
\newpage{}

\section{Introduction}\label{Sec:intro}

The rapidly advancing sensitivity of gravitational wave observatories is opening a new discovery channel for particle physics and cosmology.
In particular, stochastic gravitational wave backgrounds sourced by first-order phase transitions
in the early Universe provide a compelling probe of hidden sectors, extended gauge symmetries,
and early Universe scalar dynamics.
First-order phase transitions generate gravitational waves with a spectral shape primarily
controlled by a small set of macroscopic transition parameters such as the strength parameter $\alpha$,
the inverse duration $\beta/H_\ast$, and the characteristic temperature scale $T_\ast$.
These quantities are determined by the underlying microscopic model parameters, including the scalar vacuum expectation values (vevs), scalar quartic couplings, and gauge couplings. As a result, gravitational wave detectors can probe the microscopic properties of the scalar and gauge sectors responsible for the phase transition.
Crucially, current pulsar timing arrays (PTAs)~\cite{NANOGrav:2023gor,Reardon:2023gzh,EPTA:2023fyk} and upcoming space-based interferometers, such as Taiji~\cite{Ruan:2018tsw}, TianQin~\cite{TianQin:2015yph}, and LISA~\cite{LISA:2017pwj}, together with proposed next-generation ground-based experiments such as CE~\cite{LIGOScientific:2016wof} and ET~\cite{Punturo:2010zz}, are expected to probe a wide class of beyond the Standard Model (SM) scenarios. Their complementary frequency sensitivities, ranging from the nanoHz band to the Hz--kHz band, make them sensitive to phase transitions occurring over a broad temperature range, corresponding to scalar vevs from the MeV scale up to the PeV scale.

From a phenomenological standpoint, however, turning this promise into
concrete predictions and experimentally relevant benchmarks remains challenging.
The theoretical pipeline from a microscopic particle physics model to an observable gravitational wave spectrum
is typically computationally intensive and numerically delicate:
one must evaluate an effective potential of a Higgs-like scalar field,
determine nucleation and percolation conditions, solve for key thermodynamic quantities,
and propagate these to the gravitational wave spectrum with appropriate redshifting and detector response.
When exploring multi-dimensional parameter spaces,
including couplings, masses, vevs, etc.,
this cost is amplified by the large number of parameter points required for a robust scan.
Traditional Monte Carlo (MC) approaches sample the parameter space from a fixed, preassigned distribution, often taken to be uniform or log-uniform over the specified parameter ranges, and then retain points that satisfy the required physical-regime validity criteria and experimental constraints. Among the retained points, however, the MC sampling is not adaptive to the physical goal of finding large gravitational wave amplitudes in detector sensitive frequency regions. As a result, it may spend substantial computational effort on viable but detector-irrelevant regions of parameter space. In general, MC scans are straightforward and statistically well defined when the goal is to generate a representative background sample or to estimate global properties, such as the fraction of viable parameter space or the distributions of derived quantities. However, they are highly inefficient for the practical task of \emph{discovering viable and detectable gravitational wave signals}.

This inefficiency arises from a central feature of gravitational wave phenomenology: viable and detectable benchmark points typically occupy only a small fraction of the total parameter space scanned for a given model.
A large fraction of parameter points may yield a mathematically ``successful'' phase transition, yet remain of limited phenomenological relevance for observational studies: they may be physically inconsistent and must therefore be discarded, or they may generate gravitational wave spectra that are too weak or peak outside the sensitive frequency ranges of current and planned detectors.
In other words, if the goal is not to statistically characterize all viable scanned points, but rather to identify targeted events, such as detectable gravitational wave signals, then an unguided MC scan spends most of its evaluations on points that are uninformative for current or near-future experimental prospects. This situation reflects a broader theme in modern scientific computing: many frontiers in physics are increasingly limited not by conceptual modeling itself, but by the efficient navigation of large hypothesis spaces under computationally intensive simulations and intricate constraints.

Modern machine learning has become a standard tool in high energy physics, with widespread applications at the LHC ranging from event reconstruction and classification to jet substructure, fast simulation, and anomaly searches~\cite{Larkoski:2017jix,Guest:2018yhq}. More recently, machine learning has also entered theory and phenomenology workflows, where a central challenge is to efficiently navigate large, structured parameter spaces under computationally intensive calculations. In this context, reinforcement learning is particularly well suited for goal-directed exploration: rather than sampling blindly, a reinforcement learning agent learns adaptive search policies from experience and identifies regions of parameter space that are more likely to yield high-value outcomes under a user-defined objective. Reinforcement learning has already been applied to discrete landscape and model-building problems, including searches for string vacua and viable flavor constructions, often achieving substantial improvements over unguided random walks or naive scans~\cite{Halverson:2019tkf,Harvey:2021oue,Nishimura:2020nre,Wojcik:2024azu,Nishimura:2024apb,Carta:2025asr,Baretz:2025zsv}.
Motivated by these developments, this work adopts reinforcement learning to accelerate the discovery of experimentally relevant gravitational wave benchmark points in high-dimensional parameter spaces, where each model evaluation is computationally intensive, highly time-consuming, and detectable signals occupy only a small fraction of the scanned region.

In our application, the objective is explicitly detection-driven:
we seek parameter points that
\emph{(1) realize a successful and physically consistent first-order phase transition and
(2) produce a gravitational wave spectrum with sufficiently large amplitude and a peak frequency within the sensitive frequency range of a target detector.}
Such strong stochastic signals are typically associated with supercooled phase transitions occurring in the low-temperature regime, where the transition temperature lies well below the masses of the heavy modes integrated out of the effective theory.
Supercooled phase transitions have therefore attracted substantial recent attention because of their relatively large gravitational wave signatures and their potential relevance to the PTA signal~\cite{Leitao:2015fmj,Megevand:2016lpr,Kobakhidze:2017mru,Iso:2017uuu,Ellis:2019oqb,Wang:2020jrd,Ellis:2020nnr,Nakai:2020oit,Addazi:2020zcj,Kawana:2022fum,Freese:2022qrl,Lewicki:2022pdb,Kierkla:2022odc,Sagunski:2023ynd,Madge:2023dxc,Fujikura:2023lkn,Athron:2023mer,Ghosh:2023aum,Athron:2023rfq,Goncalves:2024lrk,Athron:2025pog,Costa:2025csj,Li:2025nja,Feng:2025wvc,Wang:2026wwp,Pascoli:2026tuu,Salvio:2026bco,Feng:2026iqo,Xie:2026vor,Li:2026uhm}.

However, a central obstacle to making reliable gravitational wave predictions in gauge theories is the gauge dependence of the finite-temperature effective potential and the associated tunneling action. The Higgs effective potential as a function of the background field is not itself a physical observable. In the $R_\xi$ gauge, it depends explicitly on the gauge-fixing parameter $\xi$, and a conventional computation based directly on the gauge-dependent effective potential can therefore lead to gauge-dependent gravitational wave results, which can hardly be regarded as genuine physical predictions. To address this issue, we adopt a gauge-independent formulation of the effective action based on the Nielsen identity~\cite{Metaxas:1995ab,Patel:2011th,Garny:2012cg,Andreassen:2014eha,Andreassen:2014gha,Arunasalam:2021zrs,Lofgren:2021ogg,Hirvonen:2021zej,Zhu:2025pht,Liu:2025ipj,Liu:2026ask,Feng:2025wvc,Feng:2026iqo}, focusing in particular on low-temperature phase transitions relevant for experimental detection~\cite{Feng:2025wvc,Feng:2026iqo}. In our gauge-independent low-temperature scan, the physically consistent criterion is therefore that \emph{(3) the transition occurs within the low-temperature regime, with the percolation temperature satisfying $T_p/(g_x v_x)\lesssim 0.5$.}

We cast the parameter exploration task as a sequential decision process, in which an agent proposes updates to model parameters and receives a reward determined by the resulting gravitational wave observables and the relevant physical constraints. We implement this strategy using \emph{Proximal Policy Optimization} (PPO)~\cite{Schulman:2017epl}, a widely used policy-gradient reinforcement learning algorithm known for its stable performance in continuous action spaces. PPO improves training stability by updating a stochastic policy while explicitly constraining each update to remain close to the previous policy, thereby alleviating the instabilities that can arise in naive policy-gradient methods.

The remainder of this paper is organized as follows. In Section~\ref{Sec:DarkU1}, we review gravitational waves from phase transitions in a minimal dark $U(1)_x$ sector. We summarize the finite-temperature effective potential, the gauge-independent low-temperature formulation based on the Nielsen identity, and the computation of the gravitational wave spectrum. In Section~\ref{Sec:PPOGW}, we introduce the PPO algorithm and describe how it is adapted to gravitational wave parameter space exploration. We then present four different PPO reward designs used in this work, including the general scan, detector target scan, history-informed global scan, and fixed boundary scan. In Section~\ref{Sec:Results}, we compare the PPO scans with conventional MC scans in both narrow and broad vev windows, and discuss the performance of policy scale transfer. Finally, Section~\ref{Sec:Con} summarizes our conclusions and outlines possible future applications. PPO notation and terminology are summarized in Appendix~\ref{App:PPO}.

\section{Gravitational waves from a minimal dark $U(1)$ sector}\label{Sec:DarkU1}

Since the SM exhibits only a smooth crossover at the electroweak scale, hidden sectors such as a minimal dark $U(1)_x$ sector provide well-motivated frameworks for realizing first-order phase transitions and the associated gravitational wave signals.
In light of the continuing non-observation of dark matter in conventional experimental searches, gravitational waves from dark symmetry breaking provide an important multi-messenger probe of new physics in the dark sector.
In this section, we briefly review phase transitions in the minimal dark $U(1)_x$ sector and the gauge-independent formulation of gravitational wave predictions in the low-temperature regime,
which is particularly relevant for experimental detection studies.

Using the formulation reviewed in this section, we develop a computational pipeline for the minimal dark $U(1)_x$ sector that identifies parameter sets $(v_x,g_x,\lambda_x)$ realizing a successful first-order phase transition and computes the peak frequency and peak amplitude of the resulting gravitational wave spectrum as
\begin{equation}
\left(
f_{\rm GW},
\Omega_{\rm GW} h^2
\right)
= F_{\rm GW}(v_x, g_x, \lambda_x)\,,\label{eq:FGW}
\end{equation}
where $F_{\rm GW}$ denotes the computational machinery for the full phase transition and gravitational wave calculation.

\subsection{Finite-temperature effective potential in the minimal $U(1)_x$ sector}

In this subsection, we briefly review the minimal $U(1)_x$ extension of the SM
and present the $U(1)_x$ Higgs effective potential relevant for gravitational wave computations in the $R_\xi$ gauge.

The minimal $U(1)_x$ extension of the SM consists of
a dark Higgs, a dark photon, and optionally dark fermion dark matter candidates.
The full Lagrangian can be written as
\begin{equation}
\mathcal{L}=\mathcal{L}_{{\rm SM}}+\mathcal{L}_{{\rm hid}}+\mathcal{L}_{{\rm mix}}\,,
\end{equation}
where $\mathcal{L}_{{\rm SM}}$ and $\mathcal{L}_{{\rm hid}}$ are
given by
\begin{align}
\mathcal{L}_{{\rm SM}}\supset & -\frac{1}{4}F_{\mu\nu}F^{\mu\nu}-\mu_{{\rm SM}}^{2}H^{\dagger}H+\lambda_{{\rm SM}}(H^{\dagger}H)^{2}\,,\\
\mathcal{L}_{\mathrm{hid}}= & \;-\frac{1}{4}F_{x\,\mu\nu}F_{x}^{\mu\nu}+\bigl|(\partial_{\mu}-ig_{x}A_{\mu}^{\prime})\Phi\bigr|^{2}+\mu_{x}^{2}\Phi^{\ast}\Phi-\lambda_{x}(\Phi^{\ast}\Phi)^{2}\nonumber \\
 & +g_{x}\overline{\chi}\gamma^{\mu}\chi A_{\mu}^{\prime}+\overline{\chi}(i\gamma^{\mu}\partial_{\mu}-m_{\chi})\chi\,,\label{eq: LintSY}
\end{align}
where $F_{\mu\nu},F_{x\,\mu\nu}$ are the field strengths of the hypercharge
and $U(1)_{x}$ gauge fields, respectively, $H$ is the SM
Higgs doublet and $\Phi$ is the $U(1)_{x}$ Higgs field with $U(1)_{x}$
charge $+1$. $A_{\mu}^{\prime}$ is the $U(1)_{x}$ gauge field,
and $\chi$ is a dark fermion with vector mass $m_{\chi}$ and $U(1)_{x}$
charge $Q_{x}=+1$. The most important parameters relevant for the
study of both $U(1)_{x}$ gravitational wave physics and dark matter
physics are the $U(1)_{x}$ gauge coupling $g_{x}$, the $U(1)_{x}$
Higgs vev $v_{x}$ and its quartic coupling $\lambda_{x}$.

The mixing terms contain kinetic mixing and Higgs mixing contributions,
written as
\begin{equation}
\mathcal{L}_{{\rm mix}}=\;-\frac{\delta}{2}F_{\mu\nu}F_{x}^{\mu\nu}-\lambda_{{\rm mix}}(H^{\dagger}H)(\Phi^{\ast}\Phi)\,,\label{eq:KM}
\end{equation}
where $\delta$ and $\lambda_{{\rm mix}}$ are mixing parameters,
both receiving stringent constraints from experiments.

After expanding $\Phi$ around the background field $\phi_{c}$ as
$\Phi=\frac{1}{\sqrt{2}}\bigl(\phi_{c}+h_{x}+iG\bigr)$, where $h_{x}$ and $G$ are the dark Higgs and Goldstone boson, respectively,
the full effective potential in the $R_\xi$ gauge is given by
\begin{equation}
V_{{\rm eff}}(\phi_{c},T,\xi)=V_{0}^{\rm tree}(\phi_{c})+V_{0}^{{\rm \text{1-loop}}}(\phi_{c},\xi)+V_{T}^{{\rm {\rm \text{1-loop}}}}(\phi_{c},T,\xi)+V_{{\rm daisy}}(\phi_{c},T,\xi)\,, \label{eq:VFull}
\end{equation}
where $V_{0}^{\rm tree}$ is the tree-level potential, $V_{0}^{{\rm {\rm \text{1-loop}}}}$
and $V_{T}^{{\rm {\rm \text{1-loop}}}}$ represent the one-loop
corrections at zero and finite temperatures respectively, and $V_{{\rm daisy}}$
corresponds to the daisy contribution arising from higher-loop resummation.

In the conventional gravitational wave analysis,
one starts from the full gauge-dependent effective potential in Eq.~(\ref{eq:VFull}),
and solves the bounce equation Eq.~(\ref{eq:S3EoM}) in the Landau gauge $\xi=0$.
This leads to a gauge-dependent bounce solution, and consequently, gauge-dependent gravitational wave results,
preventing definitive predictions for the minimal dark $U(1)_x$ sector.

\subsection{Gauge-independent formulation for low-temperature phase transitions}

\subsubsection{Gauge-independent effective action}

In gauge theories, the effective potential evaluated at a fixed background field value is an off-shell quantity and is therefore not a physical observable. In the $R_\xi$ gauge, $V_{\rm eff}(\phi_c,T,\xi)$ generally depends explicitly on the gauge-fixing parameter $\xi$.
Physical observables derived from the full effective action are gauge-independent when computed consistently, but practical phase transition calculations rely on truncated derivative and loop expansions. As a result, quantities extracted directly from a truncated gauge-dependent potential in a fixed gauge such as the Landau gauge, may retain residual gauge dependence. Since the nucleation rate and the resulting gravitational wave spectrum depend on the tunneling action, a gauge-independent prediction requires a consistent organization of the effective action rather than a direct use of the gauge-dependent potential with a chosen gauge-fixing parameter $\xi$.

The gauge dependence of the effective action is controlled by the Nielsen identity~\cite{Metaxas:1995ab},
\begin{equation}
\xi\frac{\partial S_{{\rm eff}}}{\partial\xi}=-\int{\rm d}^{4}x\,\frac{\delta S_{{\rm eff}}}{\delta\phi_{c}(x)}\,C(x)\,,\label{eq:NI}
\end{equation}
where $\phi_{c}$ denotes the background field and $C(x)$ is the Nielsen function.
This identity shows that the gauge variation of the effective action can be compensated by a field redefinition of the background field.
Therefore, when the action is evaluated on configurations satisfying the equations of motion,
the gauge dependence cancels order by order in a consistent expansion.

For phase transition applications, we use a derivative expansion of the effective action
truncated at second order in derivatives
\begin{equation}
S_{{\rm eff}} =\int{\rm d}^{4}x\,\bigl[V_{{\rm eff}}(\phi_{c})+\tfrac{1}{2}Z(\phi_{c})(\partial_{\mu}\phi_{c})^{2}
+\mathcal{O}(\partial^{4})\bigr]\,,\label{eq:GradSeff}\\
\end{equation}
where $Z$ is the field strength renormalization and the finite-temperature effective potential $V_{{\rm eff}}(\phi_{c})$
is usually computed up to one- or two-loop order.
The Nielsen function is also expanded as $C(x) = C_0(x) + \mathcal{O}(\partial^2)$ where $C_0(x)$ is the Nielsen coefficient.

The formal loop expansion must then be reorganized according to the appropriate power counting in the thermal regime of interest. In a weakly coupled $U(1)_x$ theory, loop corrections are organized by powers of $g_x^2$ and $\lambda_x$, and the relation $\lambda_x \sim g_x^n$ depends on whether the transition is treated in the high- or low-temperature regime. The gauge coupling power counting then determines which terms belong to the leading and subleading effective actions and is essential for maintaining gauge independence through the Nielsen identity.

We decompose the effective potential and the field strength renormalization
schematically as
\begin{equation}
V_{\rm eff}
=V_{\rm LO}+V_{\rm NLO}+\cdots\,,
\qquad
Z=1+Z_{\rm NLO}+\cdots\,.
\label{eq:LO_NLO_decomposition}
\end{equation}
The leading-order potential is chosen to be \emph{explicitly gauge-independent} and is used to determine the bounce configuration,
\begin{equation}
\nabla^2\phi_b
=\left. \frac{\partial V_{\rm LO}}{\partial \phi_c} \right|_{\phi_c=\phi_b}\,.
\end{equation}
The subleading tunneling action is then evaluated on this leading-order bounce,
\begin{equation}
S_{\rm NLO}[\phi_b]
=
\int {\rm d}^3x
\left[
V_{\rm NLO}(\phi_b)
+
\tfrac{1}{2}Z_{\rm NLO}(\phi_b)(\partial_\mu \phi_b)^2
\right] .
\label{eq:SNLO_compact}
\end{equation}
The correction to the bounce profile induced by $S_{\rm NLO}$ does not contribute to the tunneling action at this order~\cite{Feng:2026iqo}.

The relevant Nielsen identities at this order take the schematic form
\begin{equation}
\xi\frac{\partial V_{\rm LO}}{\partial \xi}
=0\,,
\qquad
\xi\frac{\partial V_{\rm NLO}}{\partial \xi}
=-C_{\rm LO} \frac{\partial V_{\rm LO}}{\partial \phi_c}\,,
\qquad
\xi\frac{\partial Z_{\rm NLO}}{\partial \xi}
=-2\frac{\partial C_{\rm LO}}{\partial \phi_c}\,.
\end{equation}
Using these relations, the subleading action can be verified to be independent of the gauge-fixing parameter $\xi$.
Thus, the leading-order action is manifestly gauge-independent, and the subleading action is gauge-independent by virtue of the Nielsen identity. Consequently, the nucleation rate and phase transition quantities obtained from the consistently truncated effective action are gauge-independent to the order considered.

In this work, we therefore adopt the following prescription. For each temperature regime, we first apply the appropriate power counting, decompose the effective action into leading and subleading parts, and solve for the bounce using the $\xi$-independent leading-order potential. We then evaluate the subleading action on this bounce. The nucleation quantities and the resulting gravitational wave spectrum obtained in this way are thus gauge-independent up to the order retained in the controlled expansion.

\subsubsection{Low-temperature phase transitions}

As discussed in detail in~\cite{Feng:2026iqo}, high-temperature phase transitions generally produce weaker gravitational wave signals and are therefore phenomenologically less relevant for planned space-based detectors. In this paper, we therefore focus on low-temperature phase transitions, which can generate stronger gravitational wave signals and are thus  more promising from the perspective of detection.

The low-temperature regime is defined by the hierarchy between the transition temperature and the characteristic mass scale of the dark sector, $T_p\ll m_{A'}(v_x)$. This criterion is distinct from supercooling, which instead characterizes the dynamical delay of the transition, typically through $T_p/T_c\ll1$. In practice, low-temperature transitions are naturally associated with supercooling. Indeed, benchmark points satisfying the low-temperature criterion $T_p/m_{A'}\lesssim0.2$ can be regarded as at least moderately supercooled phase transitions~\cite{Feng:2026iqo}.

In this regime, the thermal corrections are exponentially suppressed in the broken-phase region, where $\phi_c\sim v_x$ and $m(\phi_c)/T\gg1$,
\begin{equation}
\frac{T^{4}}{2\pi^{2}}J_B\left(\frac{m^2(\phi_c)}{T^2}\right)
\simeq
-\frac{T^{4}}{2\pi^{2}}\sqrt{\frac{\pi}{2}}
\left(\frac{m(\phi_c)}{T}\right)^{3/2}
\exp\left(-\frac{m(\phi_c)}{T}\right)\,.
\end{equation}
However, they become significant near the origin, where $m_{A'}(\phi)=g_x\phi\lesssim T_p$.
In this small-field region, the dark gauge boson thermal contribution behaves as
\begin{equation}
V_T^{A'}(\phi,T)
=
\frac{3T^4}{2\pi^2}
J_B\left(\frac{g_x^2\phi^2}{T^2}\right)
\simeq
-\frac{\pi^2T^4}{30}
+\frac{g_x^2T^2}{8}\phi^2
-\frac{g_x^3T}{4\pi}\phi^3+\cdots .
\end{equation}
The positive thermal mass contribution can lift the tachyonic Higgs mass term at the origin and render the origin a local minimum.
At the same time, the cubic term, together with the daisy contribution, generates a shallow barrier near the origin.
Thus a successful low-temperature first-order transition requires a parametrically small tachyonic mass term and a loop-sized quartic coupling, such that the thermal barrier can compete with the tree-level quartic term. We therefore use the low-temperature power counting
\begin{equation}
\lambda_x\sim g_x^4,
\qquad
\mu_x^2\sim g_x^4\phi_c^2 .
\label{eq:counting_LT}
\end{equation}
With this scaling, the characteristic momentum of the bounce is $k_{\rm field}\sim \sqrt{\lambda_x}\phi_{\rm wall}\sim g_x^2\phi_{\rm wall}$, while the heavy scale is $\Lambda\sim g_x\phi_{\rm wall}$. Hence $k_{\rm field}/\Lambda\sim g_x\ll1$, and the derivative expansion remains parametrically controlled.

Following this power counting, the leading and subleading low-temperature effective potentials are organized as
\begin{align}
V^{\rm LT}_{g_{x}^{4}} & = -\frac{\mu_{x}^{2}}{2}\phi_{c}^{2}+\frac{\lambda_{x}}{4}\phi_{c}^{4}+3\frac{m_{A^{\prime}}^{4}}{64\pi^{2}}\Bigl(\log\frac{m_{A^{\prime}}^{2}}{\Lambda^{2}}-\frac{5}{6}\Bigr) \nonumber\\
&\quad + V_{{\rm daisy}}^{A^{\prime}}(\phi,T) 
+\frac{T^{4}}{2\pi^{2}}\left[J_{B}\Bigl(\frac{m_{h_{x}}^{2}}{T^{2}}\Bigl)+3J_{B}\Bigl(\frac{m_{A^{\prime}}^{2}}{T^{2}}\Bigl)\right]\,,\\
V^{\rm LT}_{g_{x}^{6}} & = \frac{m_{G}^{4}}{64\pi^{2}}\Bigl(\log\frac{m_{G}^{2}}{\Lambda^{2}}-\frac{3}{2}\Bigr)-\frac{m_{c}^{4}}{64\pi^{2}}\Bigl(\log\frac{m_{c}^{2}}{\Lambda^{2}}-\frac{3}{2}\Bigr) \nonumber\\
& \quad +	V_{{\rm daisy}}^{G}(\phi,T)
+\frac{T^{4}}{2\pi^{2}}\left[J_{B}\Bigl(\frac{m_{G}^{2}}{T^{2}}\Bigr)-J_{B}\Bigl(\frac{m_{c}^{2}}{T^{2}}\Bigr)\right]\,,
\\
%
Z &=1+Z_{g_{x}^{2}}+\delta Z_{g_{x}^{2}}^T+\mathcal{O}(g_{x}^{4})\nonumber\\
&\approx 1+ \frac{g_{x}^{2}}{16\pi^{2}}\Bigl(\xi\log\frac{m_{c}^{2}}{\Lambda^{2}}+3\log\frac{m_{A^{\prime}}^{2}}{\Lambda^{2}}+\xi\Bigr)
+\mathcal{O}(g_{x}^{4})\,,\\
C_{0} & =C_{g_{x}^{2}}+\mathcal{O}(g_{x}^{4}) = -\frac{\xi g_{x}^{2}\phi_{c}}{32\pi^{2}}\,\log\frac{m_{c}^{2}}{\Lambda^{2}}+\mathcal{O}(g_{x}^{4})\,.
\end{align}
The $\xi$-independent terms are included in the leading-order potential, while the $\xi$-dependent Goldstone and ghost contributions enter the subleading action. This ordering is justified both by the large-field Boltzmann suppression and by the small-field expansion near the origin, where the $\xi$-independent thermal terms appear at lower order than the $\xi$-dependent terms.

The bounce is obtained from the gauge-independent leading-order potential,
\begin{equation}
	\nabla^{2}\phi_{b}=\frac{\partial V^{\rm LT}_{g_{x}^{4}}}{\partial\phi_{c}}\biggr|_{\phi_{c}=\phi_{b}}\,,\qquad\phi_{b}(\infty)=0\,,\qquad\phi_{b}^{\prime}(0)=0\,,
\end{equation}
and the nucleation rate is
\begin{align}
\Gamma & =\mathcal{A}\,{\rm e}^{-(S^{\rm LT}_{0}+S^{\rm LT}_{1})/T}\,,\\
S^{\rm LT}_{0} & =\int{\rm d}^{3}x\,\Bigl[V^{\rm LT}_{g_{x}^{4}}(\phi_{b})+\frac{1}{2}(\partial_{i}\phi_{b})^{2}\Bigr]\,,\\
S^{\rm LT}_{1} & =\int{\rm d}^{3}x\,\Bigl[V^{\rm LT}_{g_{x}^{6}}(\phi_{b})+\frac{1}{2}Z_{g_{x}^{2}}(\partial_{i}\phi_{b})^{2}\Bigr]\,.
\end{align}
The leading action $S^{\rm LT}_0$ is manifestly gauge-independent. The subleading action is gauge-independent by the Nielsen identities,
\begin{equation}
	\xi\frac{\partial V^{\rm LT}_{g_{x}^{4}}}{\partial\xi}  =0\,,\qquad
	\xi\frac{\partial V^{\rm LT}_{g_{x}^{6}}}{\partial\xi}  =-C_{g_{x}^{2}}\frac{\partial V^{\rm LT}_{g_{x}^{4}}}{\partial\phi_{c}}\,,\qquad
	\xi\frac{\partial Z_{g_{x}^{2}}}{\partial\xi}  =-2\frac{\partial C_{g_{x}^{2}}}{\partial\phi_{c}}\,.
\end{equation}
In practice, the second Nielsen identity is satisfied to a good approximation once the small thermal contributions are included.
As a result, the subleading effective action $S^{\rm LT}_1$, evaluated on the leading-order bounce solution, is found numerically to be approximately gauge independent.
Hence, the nucleation rate and the phase transition quantities derived from the low-temperature effective action are gauge-independent up to the order retained in the controlled expansion.

\subsection{Gravitational wave spectrum from phase transitions}

At finite temperature, the bubble nucleation rate is given by~\cite{Coleman:1977py,Callan:1977pt,Linde:1981zj,Gould:2021ccf}
\begin{equation}
	\Gamma(T)=T^{4}\biggl[\frac{S_{3}(\phi_{b},T)}{2\pi T}\biggr]^{3/2}\,{\rm e}^{-S_{3}(T)/T}\,,\label{eq:Gamma-1}
\end{equation}
where $S_{3}(T)\equiv S_{3}(\phi_{b},T)$ is the three-dimensional Euclidean action, defined as
\begin{equation}
	S_{3}(\phi,T)= 4 \pi \int_{0}^{\infty}{\rm d}r\,r^2\,\left[\frac{1}{2}\Bigl(\frac{\rm{d}\phi}{\rm{d}r}\Bigr)^{2}+\widetilde{V}_{{\rm eff}}(\phi,T)\right]\,, \label{eq:S3}
\end{equation}
with $\widetilde{V}_{{\rm eff}}(\phi,T)\equiv V_{{\rm eff}}(\phi,T)-V_{{\rm eff}}(\phi_{{\rm FV}},T)$. The bounce $\phi_{b}$ satisfies the equation of motion
\begin{equation}
	\frac{{\rm d}^{2}\phi}{{\rm d}r^{2}}+\frac{2}{r}\frac{{\rm d}\phi}{{\rm d}r}=\frac{\partial}{\partial\phi}\widetilde{V}_{{\rm eff}}(\phi,T)\,,
	\label{eq:S3EoM}
\end{equation}
subject to the boundary conditions
\begin{equation}
	\phi\,(r\to \infty) = 0\,,\qquad
	\left. \frac{{\rm d}\phi}{{\rm d}r}\right|_{r=0} =0\,.
\end{equation}
In this work the bounce solution is obtained using \textbf{CosmoTransitions}~\cite{Wainwright:2011kj}.

The Hubble rate is computed including both the radiation energy density and the vacuum energy contribution, given by
\begin{equation}
	H^{2}=\frac{\rho_{{\rm rad}}+\rho_{{\rm vac}}}{3M_{{\rm Pl}}^{2}}=\frac{1}{3M_{{\rm Pl}}^{2}}\Bigl[\frac{\pi^{2}}{30}g_{\ast}T^{4}+\Delta V_{{\rm eff}}(T)\Bigr]\,,
\end{equation}
where $\Delta V_{{\rm eff}}=V_{{\rm eff}}(\phi_{{\rm FV}},T)-V_{{\rm eff}}(\phi_{{\rm TV}},T)$,
$g_{\ast}$ is the effective number of relativistic degrees of freedom
containing both the SM and hidden sectors, and $M_{{\rm Pl}}=2.4\times10^{18}$~GeV denotes the reduced Planck mass.

The critical temperature $T_{c}$ is defined by the degeneracy condition
\begin{equation}
V_{{\rm eff}}(\phi_{{\rm TV}},T_{c})=V_{{\rm eff}}(\phi_{{\rm FV}},T_{c})\,,
\end{equation}
with a potential barrier separating the two minima. The nucleation temperature $T_n$ marks the onset of efficient bubble nucleation and is estimated when one bubble is nucleated within
one Hubble volume per Hubble time, written as
\begin{equation}
	\int_{T_{n}}^{T_{c}}\frac{\Gamma(T)}{H(T)^{4}}\frac{1}{T}\,{\rm d}T\simeq1\,.\label{eq:Tn_def2}
\end{equation}
To characterize the temperature scale associated with a delayed, supercooled phase transition,
we introduce the false vacuum fraction,
\begin{equation}
	P_{f}(T)=\exp\biggl\{-\frac{4\pi}{3}v_{w}^{3}\int_{T}^{T_{c}}{\rm d}T^{\prime}\frac{\Gamma(T^{\prime})}{T^{\prime4}H(T^{\prime})}\Bigl[\int_{T}^{T^{\prime}}\frac{{\rm d}T^{\prime\prime}}{H(T^{\prime\prime})}\Bigr]^{3}\biggr\}\,,
\end{equation}
and the percolation temperature $T_p$ is defined by $P_{f}(T_{p})\approx0.7$~\cite{Leitao:2012tx, Leitao:2015fmj}.

To ensure the phase transition completes successfully, two conditions
must be checked~\cite{Athron:2022mmm,Athron:2023mer}:
(1) the transition should have an end temperature $T_e$, defined by $P_{f}(T_e)=\varepsilon\lesssim0.01$;
(2) the physical volume of the false vacuum $\mathcal{V}_{{\rm phys}}(t)=a^{3}(t)P_{f}(t)$
decreases with time,
\begin{equation}
	\frac{{\rm d}\mathcal{V}_{{\rm phys}}}{{\rm d}t}=\mathcal{V}_{{\rm phys}}\Bigl[\frac{{\rm d}}{{\rm d}t}\ln P_{f}(t)+3H(t)\Bigr]\leq0\,,
\end{equation}
and this condition should be verified at both $T_{p}$ and $T_e$.

The characteristic length scale $L_{\ast}$ entering the gravitational wave spectrum is taken to be the mean bubble separation $R_{\ast}$.  For an approximately exponential nucleation rate, $R_{\ast}$ can be estimated as
\begin{equation}
	R_{\ast}=\frac{(8\pi)^{1/3}}{\beta}v_{w}\,, \label{eq:beta}
\end{equation}
where the inverse time scale is
\begin{equation}
	\beta=-\frac{{\rm d}}{{\rm d}t}\Bigl(\frac{S_{3}}{T}\Bigr)\biggr|_{t=t_{p}}=H(T)T\frac{{\rm d}}{{\rm d}T}\Bigl(\frac{S_{3}}{T}\Bigr)\biggr|_{T=T_{p}}\,.
\end{equation}
For strongly supercooled transitions, this exponential approximation may break down. In this case, $R_{\ast}$ should instead be directly determined from the bubble number density,
\begin{equation}
	R_{\ast}(T)\equiv\bigl[n(T)\bigr]^{-1/3}=\biggl[T^3 \int_{T}^{T_{c}}{\rm d}T^{\prime}\frac{\Gamma(T^{\prime})P_{f}(T^{\prime})}{T^{\prime4}H(T^{\prime})}\biggr]^{-1/3}\,.
\end{equation}

For a dark sector phase transition, the transition strength can be characterized by two parameters,
$\alpha_{{\rm tot}}$ and $\alpha_{h}$, defined as~\cite{Breitbach:2018ddu,Ertas:2021xeh,Bringmann:2023opz,Li:2025nja},
\begin{equation}
	\alpha_{{\rm tot}}=\frac{\Delta\bar{\theta}(T_{h,p})}{3\bigl[w_{f}^{v}(T_{v,p})+w_{f}^{h}(T_{h,p})\bigr]}\,,\qquad\alpha_{h}=\frac{\Delta\bar{\theta}(T_{h,p})}{3w_{f}^{h}(T_{h,p})}\,,
\end{equation}
where $\bar{\theta}\equiv \rho-p/{c_{s,t}^{2}}$ , and $\Delta\bar{\theta}=\bar{\theta}_{f}-\bar{\theta}_{t}$. Subscripts $f$ and $t$ refer to quantities in the false and true vacua, respectively.  Here the pressure $p$ is given by
\begin{equation}
	p  \equiv-V_{{\rm eff}}(\phi_{c},T)+\frac{\pi^{2}}{90}g_{{\rm eff}}(T)T^{4}\,,\label{eq:HDpressure}
\end{equation}
where the second term accounts for the field-independent
contribution from \textit{all relativistic particle species in thermal equilibrium with the dark thermal bath at temperature $T$}. Parameters $w_{f}^{v}$ and $w_{f}^{h}$ denote the false vacuum enthalpy densities of the visible and hidden sector, respectively, and $T_{v,p}$ and $T_{h,p}$ are their corresponding percolation temperatures. $c_{s}^{2}$ is the speed of sound.  In this work, we assume that the dark sector shares the same temperature as the visible sector, $T_h = T_v$, due to efficient interactions between the two sectors.

The total parameter $\alpha_{{\rm tot}}$ determines the amplitude of gravitational waves, and $\alpha_{h}$ controls the efficiency factor $\kappa$ which describes the fraction of vacuum
energy transformed into the bulk kinetic energy of the plasma. Assuming instantaneous reheating after percolation, the reheating temperature $T_{{\rm reh}}$ is estimated as
\begin{equation}
T_{{\rm reh}}\simeq\bigl[1+ \alpha_{\rm tot}(T_{p})\bigr]^{1/4}T_{p}\,.
\end{equation}

The bubble wall velocity is computed as $v_{w}(\alpha_h,c_{s,f}^{2},c_{s,t}^{2},\Psi)$, where $\Psi \equiv w_t/w_f$ denotes the ratio of the enthalpy densities in the true and false phases, following~\cite{Ai:2023see,Ai:2024btx}. The efficiency factor $\kappa$ is evaluated using the transition strength, the sound speeds in the two phases, and the wall velocity~\cite{Giese:2020rtr,Giese:2020znk}.

The runaway criterion is determined by
\begin{equation}
	\alpha_{h,\infty}\equiv\frac{1}{18}\frac{\sum_{i}g_{i}c_{i}\Delta m_{i}^{2}T_{h,p}^{2}}{w_{f}^{h}(T_{h,p})}\,,
\end{equation}
where $c_{i}=1\,(1/2)$ for bosons (fermions),
$g_i$ denotes the number of internal degrees of freedom of particle $i$,
and $\Delta m_{i}^{2}=m_{i,t}^{2}-m_{i,f}^{2}$, with the sum running over all species that gain a mass during the
transition. In the runaway regime ($\alpha_{h,\infty}<\alpha_{h}$), efficiency
factors are given by
\begin{equation}
	\kappa_{{\rm col}}=1-\frac{\alpha_{h,\infty}}{\alpha_{h}}\,,\quad\kappa_{{\rm sw}}=\frac{\alpha_{h,\infty}}{\alpha_{h}}\,\kappa(\alpha_{h,\infty},c_{s,f}^{2},c_{s,t}^{2},v_{w})\,,\quad\kappa_{{\rm tb}}=\epsilon\kappa_{{\rm sw}}\,,
\end{equation}
where $\epsilon$ denotes the fraction of bulk motion that is turbulent.
In the non-runaway regime ($\alpha_{h,\infty}>\alpha_{h}$), the factors are
\begin{equation}
	\kappa_{{\rm col}}=0\,,\quad\kappa_{{\rm sw}}=\kappa(\alpha_{h},c_{s,f}^{2},c_{s,t}^{2},v_{w})\,,\quad\kappa_{{\rm tb}}=\epsilon\kappa_{{\rm sw}}\,,
\end{equation}
In this work we adopt $\epsilon=0.1$ according to numerical simulations~\cite{Hindmarsh:2015qta}.

The total gravitational wave signal typically receives three primary contributions:
(1) bubble collisions of the scalar field; (2) sound waves in the
bulk plasma; (3) magneto-hydrodynamic turbulence in the plasma~\cite{Kosowsky:1991ua,Hindmarsh:2013xza,Caprini:2006jb}. Accordingly,
\begin{equation}
	h^{2}\Omega_{{\rm GW}}(f)=h^{2}\Omega_{\phi}(f)+h^{2}\Omega_{{\rm sw}}(f)+h^{2}\Omega_{{\rm tb}}(f)\,,
\end{equation}
with each component parametrized as
\begin{equation}
	h^{2}\Omega(f)=h^{2}\Omega^{{\rm peak}}\mathcal{S}(f)\,,
\end{equation}
where $\Omega^{{\rm peak}}$ is the peak amplitude and $\mathcal{S}$ the spectral-shape function. The gravitational wave power spectra observed today for the three contributions are thus given by
\begin{enumerate}
	\item bubble collision
	\begin{align}
		h^{2}\Omega_{\phi}(f) & =1.67\times10^{-5}\frac{0.11v_{w}^{3}}{0.42+v_{w}^{2}}\Bigl(\frac{100}{g_{{\rm eff}}(T_{{\rm reh}})}\Bigr)^{1/3}\Bigl(\frac{\kappa_{{\rm col}}\alpha_{{\rm tot}}}{1+\alpha_{{\rm tot}}}\Bigr)^{2}\Bigl(\frac{\beta}{H_{\ast}}\Bigr)^{-2}\mathcal{S}_{\phi}(f)\,,\\
		\mathcal{S}_{\phi}(f) & =\frac{3.8(f/f_{\phi})^{2.8}}{1+2.8(f/f_{\phi})^{3.8}}\,,\\
		f_{\phi} & =1.6\times10^{-7}\,\Bigl(\frac{g_{{\rm eff}}(T_{{\rm reh}})}{100}\Bigr)^{1/6}\Bigl(\frac{T_{{\rm reh}}}{1\,{\rm GeV}}\Bigr)\Bigl(\frac{\beta}{H_{\ast}}\Bigr)\Bigl(\frac{0.62}{1.8-0.1v_{w}+v_{w}^{2}}\Bigr)\,{\rm Hz}\,.
	\end{align}
	\item sound wave
	\begin{align}
		h^{2}\Omega_{{\rm sw}}(f) & =2.65\times10^{-6}v_{w}\Bigl(\frac{100}{g_{{\rm eff}}(T_{{\rm reh}})}\Bigr)^{1/3}\Bigl(\frac{\kappa_{{\rm sw}}\alpha_{{\rm tot}}}{1+\alpha_{{\rm tot}}}\Bigr)^{2}\Bigl(\frac{\beta}{H_{\ast}}\Bigr)^{-1}\mathcal{S}_{{\rm sw}}(f)\,,\\
		\mathcal{S}_{{\rm sw}}(f) & =\Bigl(\frac{f}{f_{{\rm sw}}}\Bigr)^{3}\biggl[\frac{7}{4+3\bigl(f/f_{{\rm sw}}\bigr)^{2}}\biggr]^{7/2}\,,\\
		f_{{\rm sw}} & =1.9\times10^{-7}\,\frac{1}{v_{w}}\Bigl(\frac{g_{{\rm eff}}(T_{{\rm reh}})}{100}\Bigr)^{1/6}\Bigl(\frac{T_{{\rm reh}}}{1\,{\rm GeV}}\Bigr)\Bigl(\frac{\beta}{H_{\ast}}\Bigr)\,{\rm Hz}\,.
	\end{align}
	\item turbulence
	\begin{align}
		h^{2}\Omega_{{\rm tb}}(f) & =3.35\times10^{-4}v_{w}\Bigl(\frac{100}{g_{{\rm eff}}(T_{{\rm reh}})}\Bigr)^{1/3}\Bigl(\frac{\kappa_{{\rm tb}}\alpha_{{\rm tot}}}{1+\alpha_{{\rm tot}}}\Bigr)^{3/2}\Bigl(\frac{\beta}{H_{\ast}}\Bigr)^{-1}\mathcal{S}_{{\rm tb}}(f)\,,\\
		\mathcal{S}_{{\rm tb}}(f) & =\biggl(\frac{f}{f_{{\rm tb}}}\biggr)^{3}\biggl[\frac{1}{1+\bigl(f/f_{{\rm tb}}\bigr)}\biggr]^{11/3}\Bigl(1+8\pi\frac{f}{H_{\ast}^{\prime}}\Bigr)^{-1}\,,\\
		f_{{\rm tb}} & =\frac{2.7}{1.9}f_{{\rm sw}}\,,
	\end{align}
\end{enumerate}
where $f_{\phi}, f_{{\rm sw}}, f_{{\rm tb}}$ are peak frequencies,
while $H_{\ast}$ denotes the Hubble rate at gravitational wave production,
where we take $T_\ast = T_{{\rm reh}}$,
and $H_{\ast}^{\prime}$ is the redshifted Hubble rate, given by
\begin{equation}
	H_{\ast}^{\prime}=\Bigl(\frac{a}{a_{0}}\Bigr)H_{\ast}\simeq1.6\times10^{-5}\,{\rm Hz}\,\Bigl(\frac{g_{{\rm eff}}(T_{{\rm reh}})}{100}\Bigr)^{1/6}\Bigl(\frac{T_{{\rm reh}}}{100\,{\rm GeV}}\Bigr)\,.
\end{equation}

Supercooled phase transitions can thus generically generate stronger gravitational wave signals for three reasons. First, the transition strength $\alpha$ is enhanced when the radiation energy density at the transition is suppressed. Since $\rho_{\rm rad}(T_p)\propto T_p^4$, a lower percolation temperature $T_p$ typically leads to a larger $\alpha$, provided that the released vacuum energy does not decrease correspondingly. This increases the available gravitational wave sourcing fraction, $\kappa\alpha/(1+\alpha)\to\kappa$ for $\alpha\gg1$, allowing the released vacuum energy to constitute a sizable fraction of the total energy density. Second, supercooling delays the transition and can lead to larger characteristic bubble sizes. Since the duration of the transition is controlled by $\beta/H_\ast$, the characteristic length scale $R_\ast\sim v_w/\beta$ becomes larger when $\beta/H_\ast$ is smaller, enhancing the anisotropic stresses that source gravitational waves. Third, in the supercooling regime, the driving pressure can become large compared with plasma friction, and thus bubble walls may become highly relativistic or approach the runaway regime. In this case, a larger fraction of the released energy can be stored in the bubble wall and scalar field gradient energy, strengthening the collision and field-sourced contributions to the gravitational wave spectrum.
As suggested by the parametric dependence of the gravitational wave peak frequency,
\begin{equation}
f_{\rm peak} \propto T_\ast \left(\frac{\beta}{H_\ast}\right)  \times \big(\text{redshift factors}\big)\,,
\end{equation}
supercooling can also shift the signal to lower frequencies. In many cases, supercooling lowers the characteristic transition temperature $T_\ast$ and is accompanied by a smaller $\beta/H_\ast$, both of which reduce $f_{\rm peak}$. Consequently, supercooled phase transitions provide a well-motivated mechanism for generating low-frequency gravitational wave signals, including the PTA signal in the nanohertz band for sufficiently low phase transition scales.

\section{Exploring gravitational wave signatures with reinforcement learning}\label{Sec:PPOGW}

In this section, we apply the PPO algorithm to the search for detectable gravitational wave signals.
Traditionally, gravitational wave spectra from phase transitions are obtained
by scanning the model parameter space using MC sampling.
In this work, we introduce PPO as an alternative strategy for parameter space exploration.
The purpose is to use reinforcement learning to guide the scan more efficiently
toward parameter regions that yield the desired gravitational wave spectrum.

\subsection{PPO algorithm}

The PPO algorithm consists of two neural networks: a policy network and a value network. The policy network $\pi_{\theta}$ controls the behavior of the agent in the environment. Given the current state of the environment, denoted by $s_t$, the policy network outputs a probability distribution over possible actions. The agent then samples an action $a_t$ from this distribution and applies it to the environment. The second network is the value network, $V_{\phi}^{\pi_{\theta}}(s_t)$, where $\phi$ denotes its trainable parameters. This network estimates the value of the current state under the policy $\pi_{\theta}$, i.e., the expected future return starting from $s_t$ when the agent follows the policy $\pi_{\theta}$.

The reward signals obtained from the agent--environment interactions are used to update the parameters of both the policy and value networks, with the goal of maximizing the expected return over a finite horizon. In the present application, one \emph{episode} corresponds to one completed search trajectory in the environment. After a batch of rollout data, which may contain one or several episodes,
has been collected, PPO updates the value network and the policy network using the sampled trajectories stored in the rollout buffer.\footnote{The rollout buffer is a temporary storage used to collect the interaction data generated by the agent before each PPO update.
Before the rollout stage, the agent follows the old policy $\pi_{\theta_{\rm old}}$ to interact with the environment.
The resulting states, actions, rewards, value estimates, and old-policy probabilities are stored in the buffer.
Once a sufficient number of steps or episodes has been collected, these stored data are used to compute the discounted returns and advantage estimates, which are then used to update the value and policy networks.}

\emph{In this work, each episode contains 32 steps, and the rollout buffer contains 2 episodes before each update.}

The value network is trained by comparing its predicted state values with the target returns obtained from the sampled trajectories.
Specifically, the parameters $\phi$ are updated by minimizing the value network loss
\begin{equation}
L^{\rm VF}(\phi)
=\mathbb{E}_t\!
\left[
\left(
V^{\pi_{\theta}}_{\phi}(s_t)-\hat{G}_t
\right)^2
\right]\,,
\end{equation}
where $\mathbb{E}_t[\cdots]$ denotes the empirical average over the sampled time steps stored in the rollout buffer,
$V^{\pi_{\theta}}_{\phi}(s_t)$ is the value network estimate of the expected return from the current state,
and $\hat{G}_t$ denotes the \emph{estimated} discounted return constructed from the sampled trajectory,
\begin{equation}
\hat{G}_t
=\sum_{k=0}^{T-t}
\gamma^k r_{t+k}\,.
\end{equation}
Here $T$ is the length of the episode, $r_{t+k}$ is the reward obtained from the environment at step $t+k$, and $\gamma\in[0,1]$ is the discount factor, which controls the relative importance of immediate and future rewards.

In the above expressions, $\theta$ and $\phi$ collectively denote the trainable parameters of the policy and value networks, respectively.
In practice, they represent high-dimensional sets of neural network weights and biases, $\theta=\{\theta_i\}$ and $\phi=\{\phi_j\}$, which are iteratively updated during PPO training.

The policy network is updated using the PPO clipped surrogate objective.
The probability ratio is defined by
\begin{equation}
\rho_t(\theta)
=
\frac{
\pi_{\theta}(a_t|s_t)
}{
\pi_{\theta_{\rm old}}(a_t|s_t)
}\,, \label{eq:ProRatio}
\end{equation}
where $\pi_{\theta}(a_t|s_t)$ denotes the probability assigned by the policy network with parameters $\theta$ to the action $a_t$ taken in state $s_t$. In Eq.~\eqref{eq:ProRatio}, $\pi_{\theta_{\rm old}}$ denotes the old policy used to generate the sampled trajectories stored in the rollout buffer. During the rollout stage, this policy is fixed and is used to sample actions from the environment. By contrast, $\pi_{\theta}$ denotes the current policy being optimized during the update stage. Thus, the numerator gives the probability assigned by the updated policy to the same action $a_t$ in the same state $s_t$, while the denominator gives the corresponding probability under the policy that collected the trajectory. The ratio $\rho_t(\theta)$ therefore measures how much the probability of taking that action changes after the policy parameters are updated.

The PPO clipping procedure restricts this ratio, preventing the new policy from deviating too strongly from the old policy in a single update.
The PPO clipped surrogate objective is given by
\begin{equation}
L^{\rm CLIP}(\theta)
=
\mathbb{E}_t
\Big\{
\min\!
\big[
\rho_t(\theta)\hat{A}_t\,,
\operatorname{clip}\!
\big(
\rho_t(\theta),1-\epsilon,1+\epsilon
\big)\hat{A}_t
\big]
\Big\}\,.
\end{equation}
where $\mathbb{E}_t\{\cdots\}$ again denotes the empirical average over the sampled time steps in the rollout buffer. The clipping operation constrains $\rho_t(\theta)$ to the interval $[1-\epsilon,1+\epsilon]$, thereby limiting excessively large policy updates and improving training stability. In this work, we take $\epsilon=0.2$.
The \emph{estimated} advantage function is defined as
\begin{equation}
\hat{A}_t
=\hat{G}_t-
V^{\pi_{\theta}}_{\phi}(s_t)\,,
\end{equation}
which measures whether the action taken at step $t$ leads to a return larger or smaller than the value expected from the current state under the policy $\pi_{\theta}$.

Thus $\theta$ is updated by \emph{maximizing} $L^{\rm CLIP}(\theta)$,
while $\phi$ is updated by \emph{minimizing} $L^{\rm VF}(\phi)$.
By iteratively updating the parameters $\theta$ and $\phi$, the policy and value networks gradually learn an efficient exploration strategy. In particular, the policy network learns to guide the agent from random initial states in the parameter space toward regions that yield the desired $\Omega_{\rm GW}h^2$--$f_{\rm peak}$ spectrum.

\begin{figure}
\includegraphics[scale=0.5]{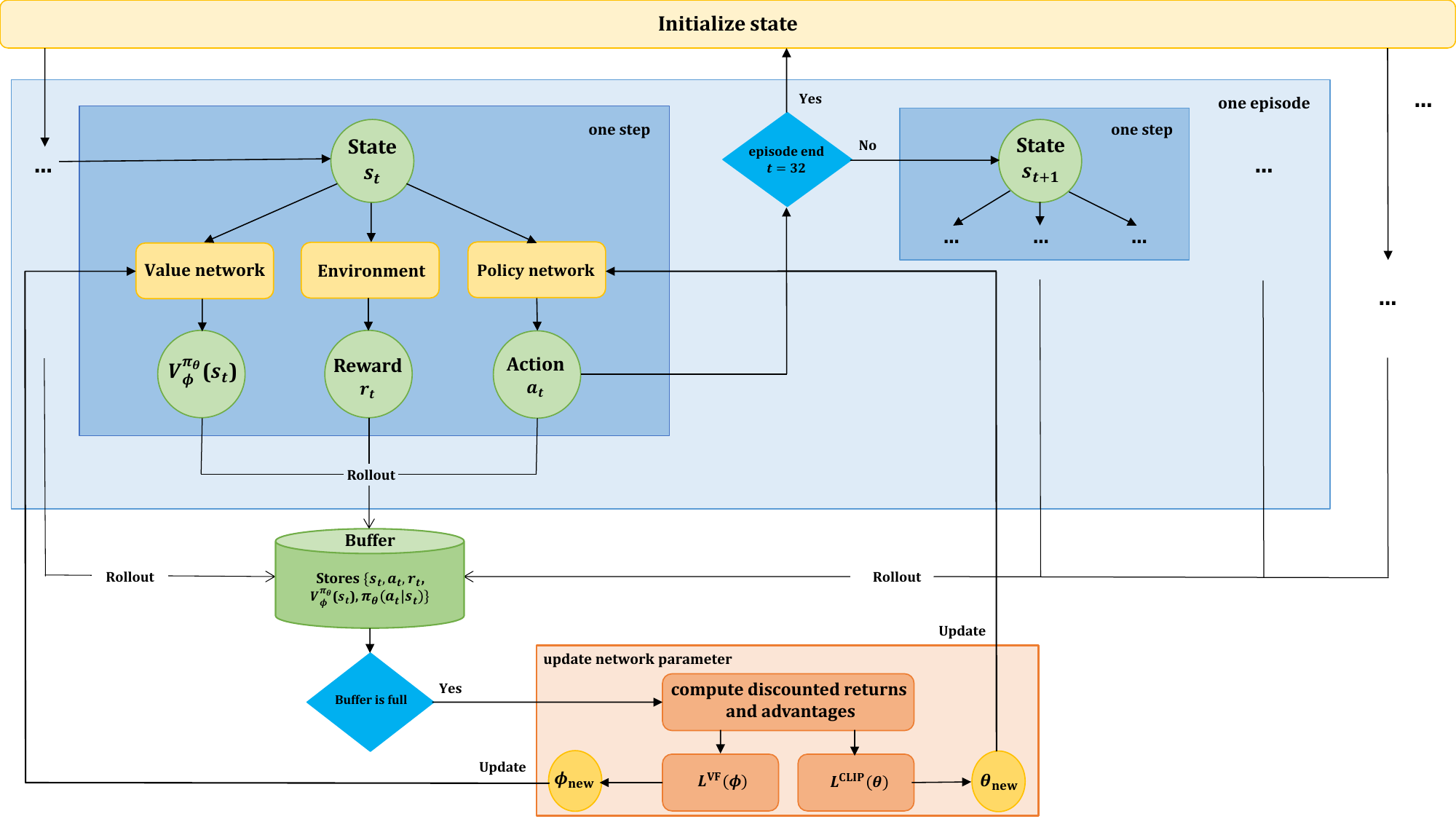}
\caption{Schematic workflow of the PPO algorithm used for parameter space exploration. After initialization, each episode consists of a sequence of rollout steps. At step $t$, the current state $s_t$ is passed to the policy network $\pi_{\theta}$, which samples an action $a_t$, and to the value network, which estimates the state value $V^{\pi_{\theta}}_{\phi}(s_t)$. The action $a_t$ is interpreted as a move in parameter space and is passed to the numerical environment, which evaluates the resulting benchmark point, computes the reward $r_t$, and constructs the next state $s_{t+1}$ for the following step.
The rollout data, including $s_t$, $a_t$, $r_t$, $V^{\pi_{\theta}}_{\phi}(s_t)$, and the action probability under the rollout policy, are stored in the buffer. After the rollout buffer is filled, the discounted returns and advantages are computed. The value network is then updated by minimizing the value function loss $L^{\rm VF}(\phi)$, while the policy network is updated using the PPO clipped surrogate objective $L^{\rm CLIP}(\theta)$. This iterative procedure allows the agent to learn an efficient exploration strategy for locating parameter regions that yield the desired $\Omega_{\rm GW}h^2$--$f_{\rm peak}$ spectrum.}
\label{Fig:PPO}
\end{figure}

The workflow of the PPO algorithm is illustrated in Fig.~\ref{Fig:PPO}. After initialization, each episode begins from an initial state in the parameter space. At a generic step $t$, the current state $s_t$ is passed to the policy network $\pi_{\theta}$, which samples an action $a_t$, and to the value network, which estimates the corresponding state value $V^{\pi_{\theta}}_{\phi}(s_t)$. In our application, the action $a_t$ is interpreted as a move in parameter space and is passed to the numerical environment. The environment evaluates the resulting benchmark point, computes the reward $r_t$ according to its ability to produce the desired gravitational wave spectrum in the $\Omega_{\rm GW}h^2$--$f_{\rm peak}$ plane, and constructs the next state $s_{t+1}$ for the following step.

The rollout tuple $\{s_t,a_t,r_t,V^{\pi_{\theta}}_{\phi}(s_t),\pi_{\theta_{\rm old}}(a_t|s_t)\}$
is stored in the rollout buffer, where $\pi_{\theta_{\rm old}}(a_t|s_t)$ denotes the probability of the sampled action under the fixed rollout policy. This procedure is repeated until the episode terminates, with one episode corresponding to one completed search trajectory in the environment. After the rollout buffer is filled, the discounted returns and advantages are computed from the stored rollout data. The value network is updated by minimizing the value function loss, while the policy network is updated using the PPO clipped surrogate objective. In particular, the stored old-policy probability $\pi_{\theta_{\rm old}}(a_t|s_t)$ is used to construct the probability ratio between the updated and old policies, thereby restricting the size of the policy update. Through repeated rollout collection and network updates, the agent gradually learns an efficient strategy for exploring the parameter space and locating regions with the desired gravitational wave signals.

\subsection{PPO for GW}

In this paper, we employ the PPO implementation provided by the Python package Stable-Baselines3~\cite{Raffin:2021stable}.
We perform four types of scans: a general scan, a history-informed global scan, a detector target scan, and a fixed boundary scan.
\begin{itemize}
  \item The \emph{general scan} is designed to identify broad regions of parameter space in which gravitational wave signals can be generated. Its reward favors points whose peak amplitudes are large relative to the best amplitude previously found in a nearby frequency region.

  \item The \emph{detector target scan} focuses on locating signals that fall within the combined sensitive ranges of proposed space-based gravitational wave experiments, and are therefore more likely to be observable.

  \item The \emph{history-informed global scan} keeps track of historical benchmark points, including the highest-amplitude points within each local frequency region, as in the general scan, as well as the lowest- and highest-frequency boundary points. It therefore encourages both broad frequency coverage and locally strong gravitational wave signals.

  \item The \emph{fixed boundary scan} uses fixed reference values for the peak frequency and peak amplitude, chosen according to prior information from preliminary scans. Its reward favors points that approach or exceed these prescribed frequency and amplitude thresholds.
\end{itemize}

\subsubsection{General scan}

In the general scan, the state at step $t$ is defined as
\begin{equation}
s_t
=
\left(
v_t\,,\,
g_t\,,\,
\lambda_t\,;\,
f_{\rm GW}^t\,,\,
\Omega_{\rm GW}^t h^2\,,\,
\Omega_{\rm ref}^t h^2
\right)\,.\label{eq:GSstate}
\end{equation}
The first three entries, $(v_t, g_t,\lambda_t)$, specify the model parameters at step $t$.
The remaining entries characterize the gravitational wave signal generated by this parameter point.
In particular
\begin{equation}
\left(
f_{\rm GW}^t,
\Omega_{\rm GW}^t h^2
\right)
= F_{\rm GW}(v_t, g_t,\lambda_t)\,,
\end{equation}
where $F_{\rm GW}$ denotes the numerical procedure used to compute the peak frequency and peak amplitude of the gravitational wave spectrum.
The action $a_t$ determines the change of the model parameters from $(v_t, g_t, \lambda_t)$ to $(v_{t+1}, g_{t+1}, \lambda_{t+1})$. The actual action is sampled from the probability distribution $\pi_\theta[a_t|s_t (v_t, g_t, \lambda_t)]$ generated by the policy network.

To reward points with large gravitational wave amplitudes while maintaining broad frequency coverage,
we introduce a dynamic reference amplitude $\Omega_{\rm ref}^t h^2$.
This quantity is not a fixed physical input, but is determined from the search history.
Specifically, it is defined as the largest gravitational wave peak amplitude previously
found in a local frequency neighborhood around the current peak frequency $f_{\rm GW}^t$.

We define the set of previously sampled gravitational wave points \emph{before} step $t$ as
\begin{equation}
\mathcal{P}_{t-1}
=
\left\{
\left(f_{\rm GW}^i,\Omega_{\rm GW}^i h^2\right)
\Big|\, i=1,\cdots,t-1
\right\}\,.
\end{equation}
The local frequency neighborhood associated with the current peak frequency $f_{\rm GW}^{t}$ is then defined by
\begin{equation}
\mathcal{N}_{t}^{f}
=\left\{
\left(f,\Omega h^2\right)\in \mathcal{P}_{t-1}
\Big|\,
\big|
\log_{10} f
-
\log_{10} f_{\rm GW}^{t}
\big| <0.1
\right\}\,.
\end{equation}
The dynamic reference amplitude is defined as the largest amplitude previously found in this local neighborhood,
\begin{equation}
\Omega_{\rm ref}^{t}h^2
=
\max_{\left(f,\Omega h^2\right)\in \mathcal{N}_{t}}
\Omega h^2\,.\label{eq:GSrefamp}
\end{equation}
If $\mathcal{N}_{t}$ is empty, an initial reference value is assigned. This dynamically defined reference amplitude is included in the state vector so that the policy and value networks have access to the local amplitude scale used in the reward function.

The reward at step $t$ is defined as
\begin{equation}
r_t
= \begin{cases}
r_t^{\Omega h^2}
= 10 \left[ 1+
\tanh\!\Big(
\dfrac{\pi}{3}
\log_{10}\!
\dfrac{\Omega_{\rm GW}^t h^2}
{\Omega_{\rm ref}^t h^2}
\Big)
\right]\,,
& \qquad \text{successful phase transition},\\[12pt]
-1\,,
& \qquad \text{no successful phase transition}.
\end{cases}\label{eq:GSRamp}
\end{equation}
This reward function assigns larger rewards to parameter points with larger gravitational wave amplitudes relative to the best signal previously found in the same local frequency region. Therefore, the general scan is not designed merely to find the globally largest amplitude; rather, it encourages the agent to identify high-amplitude signals across a broad range of peak frequencies.

After a sufficient number of steps has been collected in the rollout buffer, the policy and value networks are updated. The discounted return and advantage are computed for each step using the sampled trajectories, with the discount factor set to $\gamma=0.98$ in this work. The rollout data are then divided into mini-batches. Each mini-batch is used to update the value network parameters by gradient descent on the value loss and the policy network parameters by gradient ascent on the PPO clipped surrogate objective.

\subsubsection{Detector target scan}

In addition to the reward used in the general scan, the detector target scan introduces a detector-facing reward structure,
where the detector sensitivity curve provides a fixed reference threshold for assigning rewards.

We denote the detector sensitivity bound on the gravitational wave peak amplitude by
\begin{equation}
\Omega h^2
=
F^{\rm Bound}(f)\,.
\end{equation}
In this work, the detector target search is performed using the combined sensitivity regions of the proposed space-based gravitational wave observatories Taiji, TianQin, and LISA. We denote the frequency range covered by the detector sensitivity curve as
\begin{equation}
B_f
=\left[
f_{\rm min}^{\rm Bound}\,,
f_{\rm max}^{\rm Bound}
\right]\,,
\end{equation}
and the corresponding amplitude range as
\begin{equation}
B_{\Omega h^2}
=
\Big[
\min_{f\in B_f} F^{\rm Bound}(f)\,,
\max_{f\in B_f} F^{\rm Bound}(f)
\Big]\,.
\end{equation}
For a gravitational wave peak signal $(f_{\rm GW}^t,\Omega_{\rm GW}^t h^2)$ generated at step $t$,
the detector threshold at the same frequency is given by
\begin{equation}
\Omega_{\rm Bound}^t h^2
=
F^{\rm Bound}(f_{\rm GW}^t),
\qquad
f_{\rm GW}^t\in B_f\,.
\end{equation}

Since the detector sensitivity curve is generally not single-valued under inversion, the same gravitational wave amplitude can correspond to both the lower- and higher-frequency boundaries of the detector sensitive region.
We therefore define, for $\Omega_{\rm GW}^t h^2\in B_{\Omega h^2}$,
the set of frequencies at which the detector sensitivity curve reaches this amplitude:
\begin{equation}
S_t
=\left\{
f\in B_f\,
\Big|\,
F^{\rm Bound}(f)=\Omega_{\rm GW}^t h^2
\right\}\,.
\end{equation}
The lower and higher boundary frequencies associated with this amplitude are then
\begin{equation}
f_{\rm low}^t
=
\min S_t\,,
\qquad
f_{\rm high}^t
=
\max S_t\,.
\end{equation}
Thus, $B_f$ is used to determine whether the peak frequency lies inside the detector sensitive frequency range, while $B_{\Omega h^2}$ is used to determine whether the peak amplitude lies within the vertical range covered by the detector sensitivity region.

In the detector target scan, the state at step $t$ is again defined as
\begin{equation}
s_t
=
\left(
v_t\,,\,
g_t\,,\,
\lambda_t\,;\,
f_{\rm GW}^t\,,\,
\Omega_{\rm GW}^t h^2\,,\,
\Omega_{\rm ref}^t h^2
\right)\,,\label{eq:GSstate}
\end{equation}
in which the first three entries specify the model parameters at step $t$,
and the remaining entries characterize the gravitational wave signal generated by this parameter point.
The action $a_t$ determines the change of model parameters from $(v_t, g_t, \lambda_t)$ to $(v_{t+1}, g_{t+1}, \lambda_{t+1})$.
The actual action is sampled from the probability distribution $\pi_\theta[a_t|s_t (v_t, g_t, \lambda_t)]$ generated by the policy network.

The reward at step $t$ is defined as
\begin{equation}
r_t
=
\begin{cases}
r_t^{\Omega h^2}
+r_t^{\Omega_{\rm D} h^2}+r_t^{f_{\rm D}}\,,
& \qquad\qquad\text{successful phase transition}\,,\\[4pt]
-1\,,
& \qquad\qquad \text{no successful phase transition}\,,
\end{cases}
\end{equation}
where the amplitude reward is given by Eq.~\eqref{eq:GSRamp}.
The detector target amplitude reward is given by
\begin{equation}
r_t^{\Omega_{\rm D} h^2}
=
\begin{cases}
10\left[
1+
\tanh\!\Big(
\dfrac{\pi}{3}
\log_{10}
\dfrac{\Omega_{\rm GW}^t h^2}
{\Omega_{\rm Bound}^t h^2}
\Big)
\right]\,,
&
f_{\rm GW}^t\in B_f\,,\quad
\Omega_{\rm GW}^t h^2\leq \Omega_{\rm Bound}^t h^2\,,
\\[14pt]
10\,,
&
f_{\rm GW}^t\in B_f\,,\quad
\Omega_{\rm GW}^t h^2> \Omega_{\rm Bound}^t h^2\,,
\\[6pt]
0\,,
&
f_{\rm GW}^t\notin B_f \,.
\end{cases}
\end{equation}
This term rewards signals whose peak amplitudes approach the detector sensitivity threshold from below, while assigning the maximal reward once the signal exceeds the threshold.
The detector target frequency reward is defined as
\begin{equation}
r_t^{f_{\rm D}}
=\begin{cases}
10\left[
1+\tanh\!\Big(
\dfrac{\pi}{3}
\log_{10}
\dfrac{f_{\rm GW}^t}
{f_{\rm low}^t}
\Big)
\right]\,,
& \Omega_{\rm GW}^t h^2\in B_{\Omega h^2}\,,\quad
f_{\rm GW}^t<f_{\rm low}^t\,,
\\[14pt]
10\left[1+
\tanh\!\Big(
\dfrac{\pi}{3}
\log_{10}
\dfrac{f_{\rm high}^t}
{f_{\rm GW}^t}
\Big)
\right]\,,
& \Omega_{\rm GW}^t h^2\in B_{\Omega h^2}\,,\quad
f_{\rm GW}^t>f_{\rm high}^t\,,
\\[14pt]
10\,,
& \Omega_{\rm GW}^t h^2\in B_{\Omega h^2}\,,\quad
f_{\rm GW}^t\in
\left[
f_{\rm low}^t\,,
f_{\rm high}^t
\right]\,,
\\[8pt]
0\,,
& \Omega_{\rm GW}^t h^2\notin B_{\Omega h^2}\,.
\end{cases}
\end{equation}
This term rewards signals whose peak frequencies approach the detector sensitive region at the corresponding amplitude, and assigns the maximal reward when the peak lies inside the sensitive frequency interval.

\subsubsection{History-informed global scan}

We also introduce a history-informed global scan, which we refer to as the global scan. In this strategy, the reward is defined relative to the historical trajectory of the scan. The purpose is to encourage the agent to search for gravitational wave signals with large peak amplitudes while also improving the frequency coverage of the sampled benchmark points.

For the history-informed global scan, the state at step $t$ is defined as
\begin{equation}
s_t
=\left(
v_t\,,\,
g_t\,,\,
\lambda_t\,;\,
f_{\rm GW}^t\,,\,
\Omega_{\rm GW}^t h^2\,,\,
f_{\rm low}^{t}\,,\,
f_{\rm high}^{t}\,,\,
\Omega_{\rm ref}^t h^2
\right)\,,\label{eq:GSstate}
\end{equation}
where the corresponding gravitational wave peak frequency and amplitude $(f_{\rm GW}^t, \Omega_{\rm GW}^t h^2)$
are obtained from $F_{\rm GW}(v_t, g_t, \lambda_t)$.
The action $a_t$ determines the change of the model parameters from $(v_t, g_t, \lambda_t)$ to $(v_{t+1}, g_{t+1}, \lambda_{t+1})$. The actual action is sampled from the probability distribution $\pi_\theta[a_t|s_t (v_t, g_t, \lambda_t)]$ generated by the policy network.

The reward setup for the gravitational wave amplitude is identical to the one used in the general search.
To characterize the frequency coverage at a comparable amplitude, we also define a local amplitude neighborhood,
\begin{equation}
\mathcal{N}_{t}^{\Omega}
=\left\{
\left(f,\Omega h^2\right)\in \mathcal{P}_{t-1}\,
\Big|\,
\left|
\log_{10}(\Omega h^2)
- \log_{10}(\Omega_{\rm GW}^{t}h^2)
\right| <0.1
\right\}\,.
\end{equation}
The historically lowest and highest frequencies in this amplitude neighborhood are defined by
\begin{equation}
f_{\rm low}^{t}
= \min_{\left(f,\Omega h^2\right)\in \mathcal{N}_{t}^{\Omega}} f\,,
\qquad
f_{\rm high}^{t}
= \max_{\left(f,\Omega h^2\right)\in \mathcal{N}_{t}^{\Omega}} f\,.
\end{equation}
If either $\mathcal{N}_{t}^{f}$ or $\mathcal{N}_{t}^{\Omega}$ is empty, the corresponding reference value is assigned from an initial prescription.

The reward at step $t$ is defined as
\begin{equation}
r_t
=\begin{cases}
r_t^{\Omega h^2}+r_t^f\,,
& \qquad\qquad\text{successful phase transition},\\[4pt]
-1\,,
& \qquad\qquad\text{no successful phase transition}\,,
\end{cases}
\end{equation}
where the amplitude reward is given by Eq.~\eqref{eq:GSRamp}.
This term assigns a larger reward to points whose gravitational-wave peak amplitudes exceed the best amplitude previously found within the same local frequency neighborhood.

The frequency reward is defined as
\begin{equation}
r_t^f
=5\left[
1+
\tanh\!\Big(
\dfrac{\pi}{3}
\log_{10}
\dfrac{f_{\rm low}^{t}}
{f_{\rm GW}^{t}}
\Big)
\right] +
5\left[
1+
\tanh\!\Big(
\dfrac{\pi}{3}
\log_{10}
\dfrac{f_{\rm GW}^{t}}
{f_{\rm high}^{t}}
\Big)
\right]\,.
\end{equation}
This term rewards points that extend the historical frequency coverage at a comparable gravitational-wave amplitude, while assigning a baseline reward to points lying inside the previously covered frequency interval.

\subsubsection{Fixed boundary scan}

We also consider a fixed boundary scan, in which the reward is evaluated with respect to fixed reference values rather than history-dependent quantities. We use the minimal state defined as 
\begin{equation}
s_t
=(v_t\,,\,g_t\,,\,\lambda_t)\,,
\end{equation}
and action $a_t$ determines change of the model parameters from $(v_t,g_t,\lambda_t)$ to $(v_{t+1},g_{t+1},\lambda_{t+1})$. The actual action is sampled from the probability distribution $\pi_\theta(a_t|s_t)$ generated by the policy network.

For the fixed boundary scan,
the reference amplitude and reference frequency are fixed throughout the training process.
We choose
\begin{equation}
\Omega_{\rm FB}h^2
=
10^{-8}\,,
\qquad
f_{\rm FB}
=10^{-3}~{\rm Hz}\,.
\end{equation}
These two fixed values define the target boundary for the scan and do not change with the accumulated search history.

The reward at step $t$ is defined as
\begin{equation}
r_t
=
\begin{cases}
r_t^{\Omega h^2}+r_t^f\,,
& \qquad\qquad\text{successful phase transition},\\[4pt]
-1\,,
& \qquad\qquad\text{no successful phase transition}.
\end{cases}
\end{equation}
The amplitude reward is given by
\begin{equation}
r_t^{\Omega h^2}
=
10\left[
1+\tanh\!\Big(
\dfrac{\pi}{3}
\log_{10}
\dfrac{\Omega_{\rm GW}^{t}h^2}
{\Omega_{\rm FB}h^2}
\Big)
\right]\,.
\end{equation}
This term rewards points whose gravitational-wave peak amplitudes approach or exceed the fixed reference amplitude $\Omega_{\rm FB}h^2$.

The frequency reward is defined as
\begin{equation}
r_t^f
=
10\left[
1+
\tanh\!\Big(
\dfrac{\pi}{3}
\log_{10}
\dfrac{f_{\rm GW}^{t}}
{f_{\rm FB}}
\Big)
\right]\,.
\end{equation}
This term rewards points whose peak frequencies approach or exceed the fixed reference frequency $f_{\rm FB}$.
Therefore, this scan provides a simple fixed boundary reward prescription, in contrast to the global scan where the reference amplitude and frequency coverage are dynamically determined by the historical search trajectory.

\subsection{Important remarks}\label{sec:2Cond}

In this work, we focus on low-temperature phase transitions, which are most relevant for experimentally detectable benchmark points. We impose the low-temperature validity condition $T_p/m_{A^\prime}\lesssim 0.2$ as a \emph{selection cut}, although the low-temperature formulation can still be applied approximately in the intermediate region $0.2\lesssim T_p/m_{A^\prime}\lesssim 0.5$. On the other hand, benchmark points with $T_p/m_{A^\prime}\gtrsim 0.5$ lie outside the low-temperature regime and should be excluded.
For completeness, however, we display all scanned benchmark points in the plots, which allows us to compare the scanning efficiency of PPO and MC more transparently.

We emphasize that the low-temperature condition and the requirement of a large gravitational wave amplitude are \emph{correlated but not identical.} The condition $T_p/m_{A^\prime}\lesssim 0.2$ is a validity criterion for appropriately applying the low-temperature formulation, whereas the gravitational wave amplitude directly characterizes the detectability of the signal. In practice, benchmark points satisfying $T_p/m_{A^\prime}\lesssim 0.2$ typically correspond to supercooled phase transitions
and therefore tend to generate larger gravitational wave amplitudes, making them more relevant for experimental probes.

In the actual PPO  scan, the reward functions are designed to favor large gravitational wave amplitudes rather than to directly reward the satisfaction of the low-temperature condition $T_p/m_{A^\prime}\lesssim 0.2$. Since larger gravitational wave amplitudes are often associated with supercooling, the amplitude-based reward naturally increases the probability of finding benchmark points that satisfy the low-temperature validity condition.
Nevertheless, this distinction is important when comparing PPO with MC scans. In this work, PPO is not explicitly trained to search for points satisfying $T_p/m_{A^\prime}\lesssim 0.2$.
Therefore, using this condition as a post-selection criterion provides a conservative assessment of PPO performance,
although PPO scans already show a significant advantage over MC in narrow-window scans.

The motivation for this reward design is twofold:
\begin{itemize}
  \item From the physics perspective, our primary goal is to identify benchmark points that are most relevant for experimental detection, and therefore the reward is directly tied to the gravitational wave signal strength.
  \item From the reinforcement learning perspective, simpler reward functions generally lead to more stable and efficient training. Since large gravitational wave amplitudes and the low-temperature condition are physically correlated, we use the amplitude-based reward as a simple and effective proxy rather than introducing an additional explicit reward for $T_p/m_{A^\prime}\lesssim 0.2$.
\end{itemize}

More generally, however, the conditions required for physical consistency \emph{need not coincide} with those that optimize a particular search objective. Accordingly, the reward design in other training tasks may differ substantially from the one adopted here. One may introduce multiple reward components, with some designed to enforce physical consistency conditions and others tailored to distinct scientific objectives.

\section{Results: PPO vs MC}\label{Sec:Results}

In this work, we focus on low-temperature phase transitions in the minimal dark $U(1)_x$ sector with $v_x\in[0.1,10]~{\rm GeV}$, a range that is closely related to space-based gravitational wave detection. The PPO scan can be straightforwardly extended to other symmetry breaking scales and to other phase transition models, which we leave for future work.

\emph{To ensure a fair comparison between different scan strategies, all computations are carried out on the same laptop under the same execution setup. Each scan task is run serially on a single CPU core, without parallelization. The workstation is equipped with an AMD Ryzen AI Max+ 395 processor with Radeon 8060S graphics, running at 3.00 GHz, and 32.0 GB of installed RAM, with 23.6 GB available. The system uses a 64-bit operating system on an x64-based processor.}

We present our results for two choices of the $U(1)_x$ vev range. The first corresponds to narrow-window scans, with $v_x\in[0.1,1]~{\rm GeV}$ and $v_x\in[1,10]~{\rm GeV}$, where the scan range spans one order of magnitude. The second corresponds to a broad-window scan, with $v_x\in[0.1,10]~{\rm GeV}$, where the scan range spans two orders of magnitude. We compare the PPO scans with the MC scans from several perspectives, and provide a detailed discussion at the end of this section.

\subsection{Narrow-window scans}

In this subsection, we compare the PPO and MC scans within two narrow vev windows: $v_x\in[0.1,1]~{\rm GeV}$ and $v_x\in[1,10]~{\rm GeV}$. We find that, for the usual objective of identifying benchmark points with large gravitational wave amplitudes while maintaining broad frequency coverage, the general scan performs best among the general, history-informed global, and fixed boundary scan strategies. We therefore do not show the results of the latter two strategies in this subsection.
The detector target scan exhibits excellent performance in locating benchmark points that fall within the detector sensitive regions.

All PPO narrow-window scans are performed with \emph{80 episodes and 32 steps per episode}.

\subsubsection{Phase transitions at $v_x\in[0.1,1]~{\rm GeV}$}

We first present the results for $U(1)_x$ phase transitions in the range $v_x\in[0.1,1]~{\rm GeV}$.
The main results are summarized in Table~\ref{Tab:NWScan1}.
We compare the PPO general scan with two MC benchmarks: an MC scan performed with approximately the same computational time, denoted by MC-Time, and an MC scan that finds the same number of viable low-temperature points satisfying $T_p/m_{A^\prime}\lesssim0.2$, denoted by MC-GSPoint.
We also compare the PPO detector target scan with an MC scan that identifies the same number of potentially detectable benchmark points in the sensitive regions of space-based gravitational wave detectors. We denote this comparison sample by MC-DTPoint.

\begin{figure}[t!]
	\centering
	\subfigure[PPO-GS $\sim 8$~hours]{
	\includegraphics[scale=0.46]{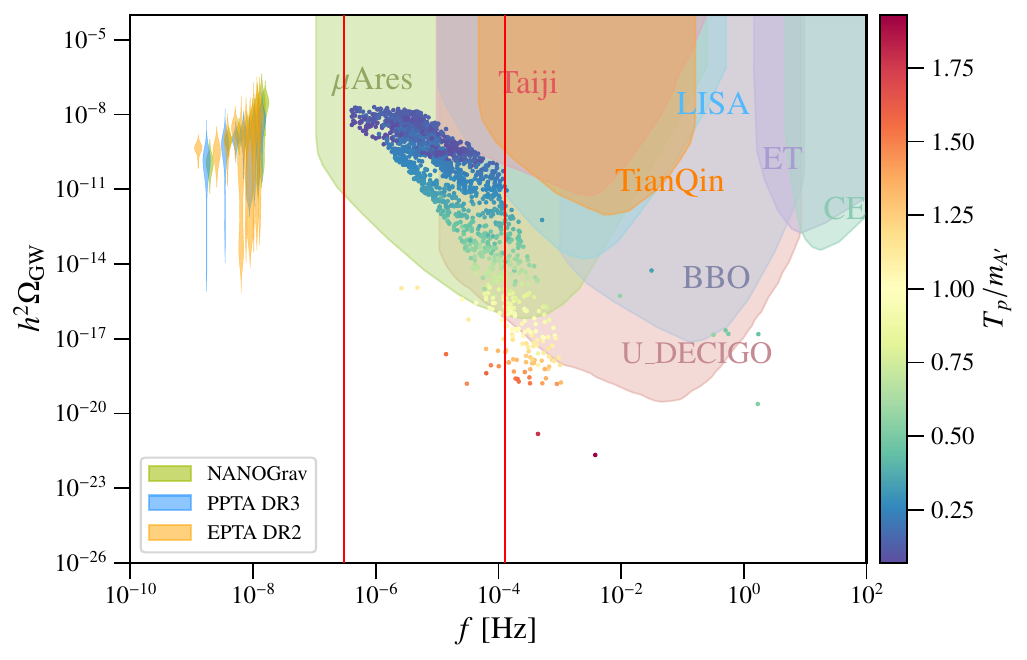}}\\
	\subfigure[MC-Time $\sim 8$~hours]{
	\includegraphics[scale=0.46]{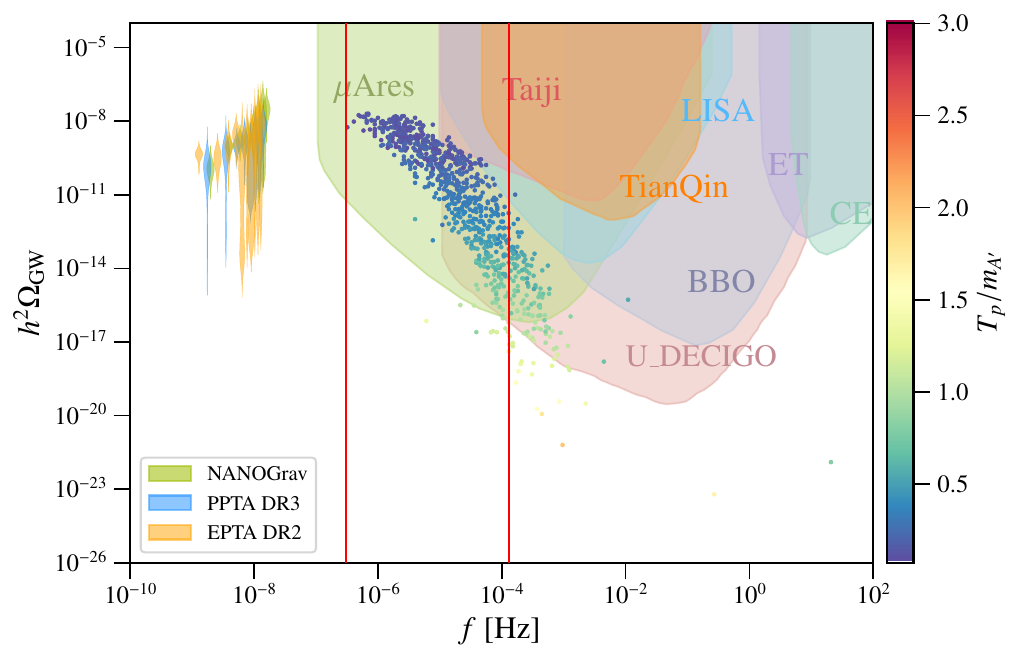}}
	\subfigure[MC-GSPoint $\sim 19$~hours]{
	\includegraphics[scale=0.46]{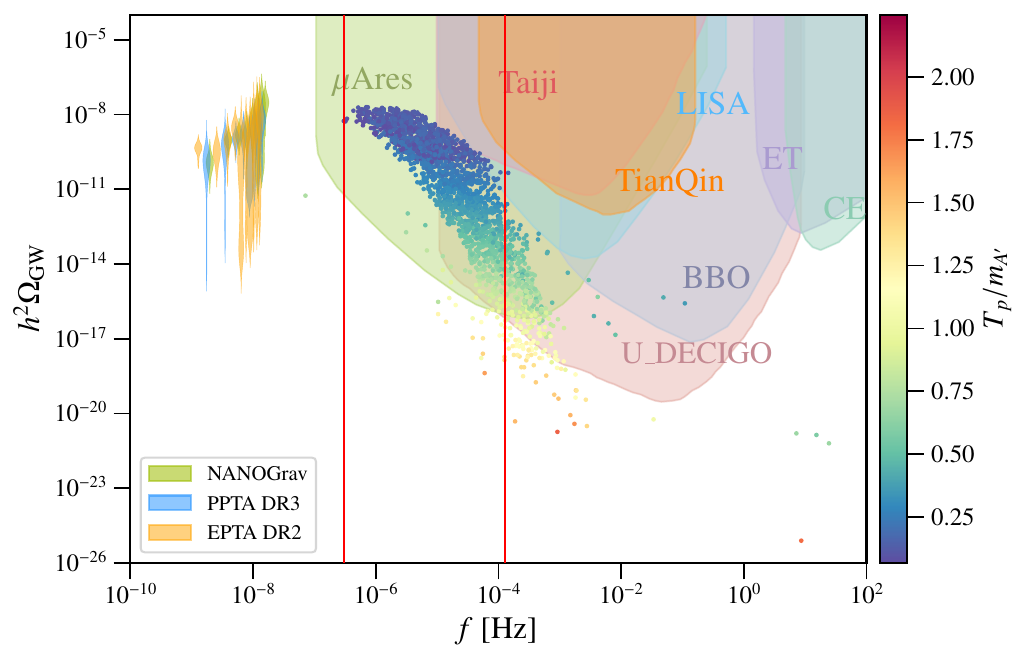}}
    \caption{Comparison of the PPO general scan with MC scans for the minimal dark $U(1)_x$ model with $v_x\in[0.1,1]~{\rm GeV}$. Panel (a) shows the PPO general scan run for about 8 hours, while panel (b) shows the MC scan performed with approximately the same runtime as PPO-GS. Panel (c) shows the MC scan performed until it obtains the same number of viable low-temperature benchmark points as PPO-GS. The two red vertical lines in each plot are the same reference lines, determined by MC-GSPoint, and indicate the frequency range reached by the MC scan after about 19 hours of runtime. The colored regions denote the power-law integrated sensitivities of current and future detectors, including Taiji~\cite{Ruan:2018tsw}, TianQin~\cite{TianQin:2015yph}, LISA~\cite{LISA:2017pwj}, $\mu$Ares~\cite{Sesana:2019vho}, BBO~\cite{Grojean:2006bp}, U$\_$DECIGO~\cite{Kuroyanagi:2014qaa}, ET~\cite{Punturo:2010zz}, and CE~\cite{LIGOScientific:2016wof}.
    The violin plots in the nanohertz region indicate the PTA signals reported by NANOGrav~\cite{NANOGrav:2023gor}, PPTA~\cite{Reardon:2023gzh}, and EPTA~\cite{EPTA:2023fyk}.}
    \label{Fig:GS1}
\end{figure}

The PPO scan exhibits a clear efficiency improvement compared with the MC scan. For approximately the same scanning time, PPO-GS finds about 2.79 times as many viable low-temperature phase transition benchmark points as the MC scan. Conversely, to find the same number of viable low-temperature benchmark points as PPO-GS, the MC scan takes about 2.43 times as much computational time. The corresponding benchmark points obtained from these scans are shown in Fig.~\ref{Fig:GS1}.
As indicated by the same red vertical reference lines, which mark the frequency range accessible to the benchmark points, the PPO-GS run of about 8 hours achieves a frequency coverage comparable to that of the much longer MC scan when searching for viable low-temperature benchmark points. By contrast, the 8-hour MC scan yields a much sparser distribution and finds far fewer viable benchmark points.

\begin{table}
\center
\begin{tabular}{|c|c|c|c|c|c|}
\hline
$0.1-1$ & Strategy & Time (s) & Total points & $T_{p}/m_{A^{\prime}}\lesssim0.2$ & Detection\tabularnewline
\hline
\hline
\multirow{2}{*}{PPO}
& PPO-GS & 27678 & 1864 & 806 & 74\tabularnewline
\cline{2-6}
& PPO-DT & 29137 & 2229 & 898 & 110\tabularnewline
\hline
\multirow{3}{*}{MC}
& MC-Time & 28567 & 1004 & 289 & 13\tabularnewline
\cline{2-6}
 & MC-GSPoint & 67182 & 2772 & 806 & 69\tabularnewline
\cline{2-6}
 & MC-DTPoint & 115431 & 5156 & 1521 & 110\tabularnewline
\hline
\end{tabular}\\
\caption{Summary of the PPO general scan and detector target scan results, together with the corresponding MC benchmarks, for the minimal dark $U(1)_x$ model with $v_x\in[0.1,1]~{\rm GeV}$.
The PPO category consists of the PPO general scan (PPO-GS) and the PPO detector target scan (PPO-DT), both performed with 80 episodes and 32 steps per episode and requiring about 8 hours of runtime. For comparison, the MC category contains three benchmark scans: MC-Time, an MC scan with approximately the same runtime as PPO; MC-GSPoint, an MC scan that finds the same number of low-temperature points satisfying $T_p/m_{A^\prime}\lesssim 0.2$; and MC-DTPoint, an MC scan required to obtain the same number of benchmark points in the sensitive regions of space-based gravitational wave detectors.
In the table, ``Total points'' denotes the total number of benchmark points for which the computation is completed and a phase transition is obtained. Among these points, ``$T_p/m_{A^\prime}\lesssim 0.2$'' counts the number of benchmark points satisfying the low-temperature validity condition, while ``Detection'' counts the number of benchmark points that fall within the space-based gravitational wave detector sensitive region.}
\label{Tab:NWScan1}
\end{table}

\begin{figure}
	\centering
	\subfigure[PPO-DT $\sim 8$~hours]{
	\includegraphics[scale=0.46]{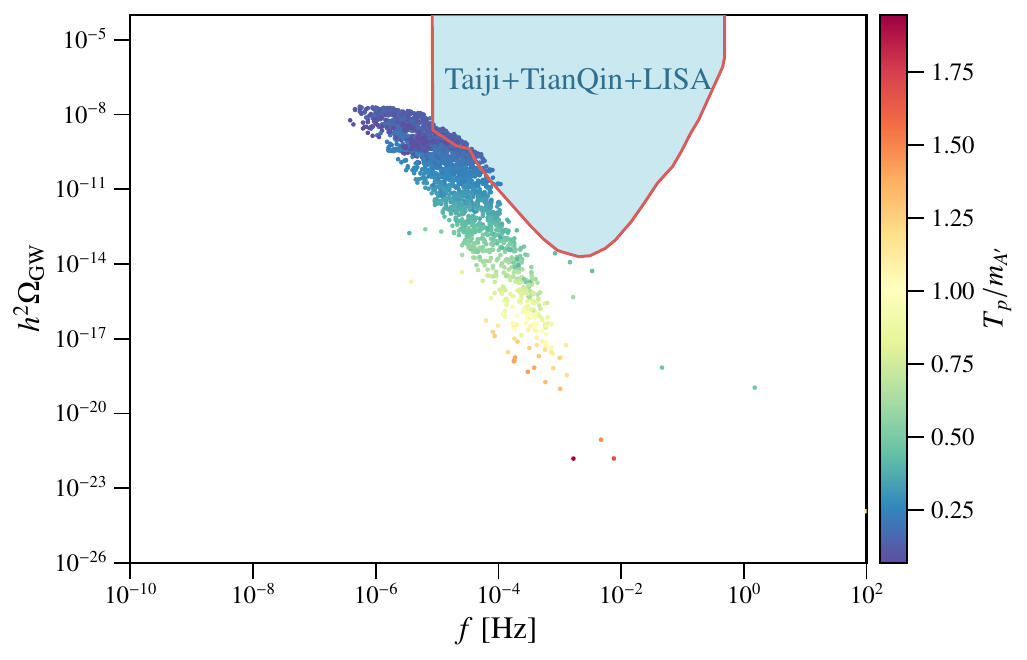}}
	\subfigure[MC2-Point $\sim 32$~hours]{
	\includegraphics[scale=0.46]{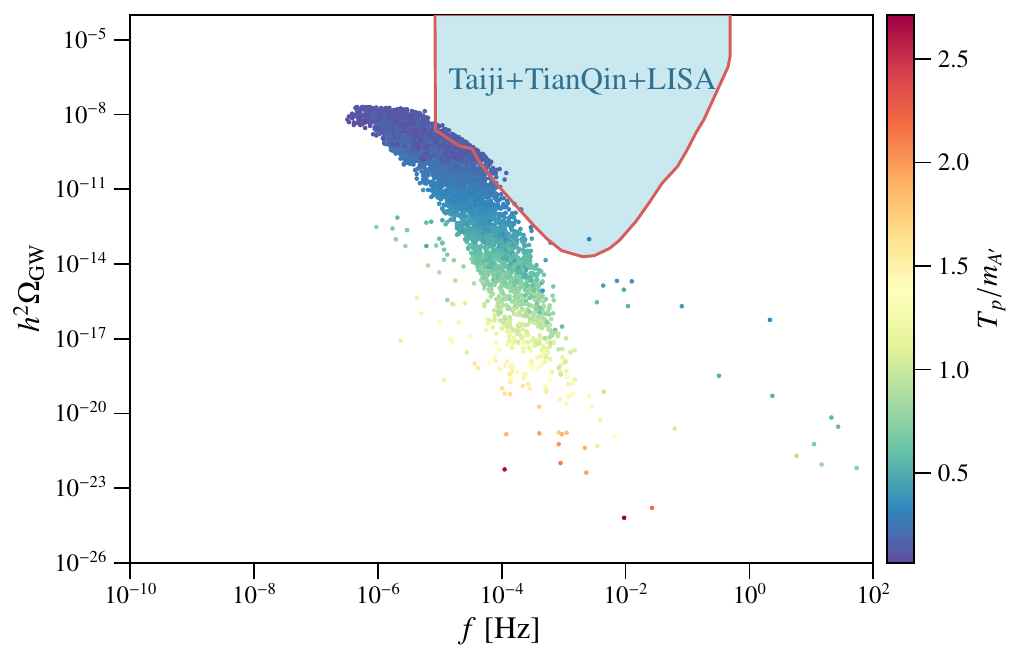}}
    \caption{Comparison of the PPO detector target scan with MC scans for the minimal dark $U(1)_x$ model with $v_x\in[0.1,1]~{\rm GeV}$. The detector target reward is designed to favor benchmark points within the combined sensitive region of Taiji, TianQin, and LISA, shown by the combined detector sensitivity curve.
    Panel (a) shows the PPO detector target scan run for about 8 hours, while panel (b) shows the MC scan performed until it obtains the same number of benchmark points in the detector sensitive region, for about 32 hours.}
    \label{Fig:DT1}
\end{figure}

\begin{figure}
	\centering
	\subfigure[PPO-GS $\sim 8$~hours]{
	\includegraphics[scale=0.54]{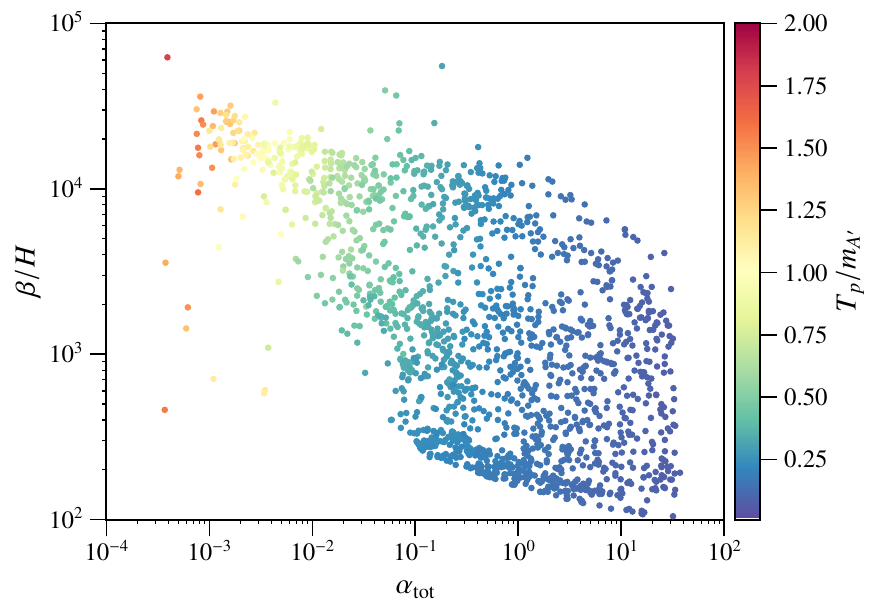}}
	\subfigure[MC-GSPoint $\sim 19$~hours]{
	\includegraphics[scale=0.54]{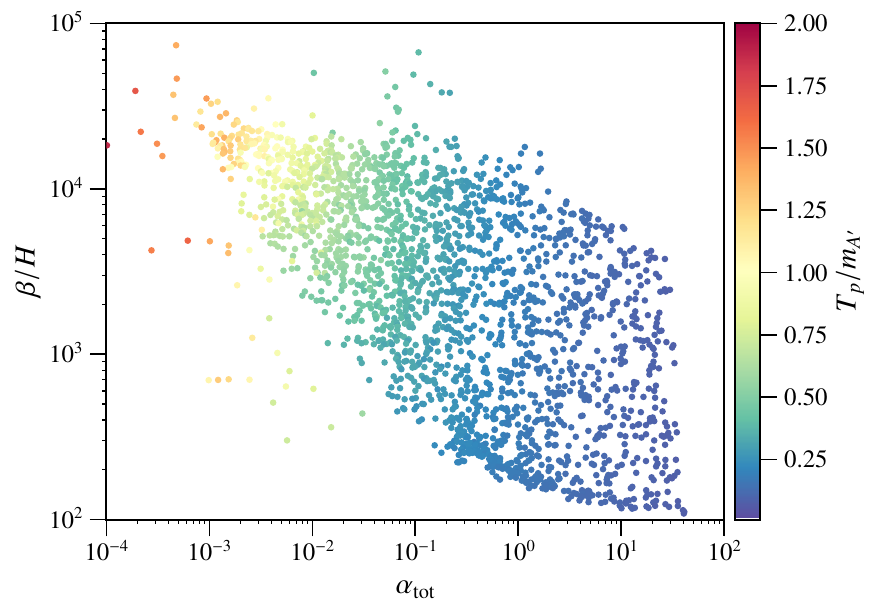}}\\
\hspace{-25pt}
	\subfigure[PPO-DT $\sim 8$~hours]{
	\includegraphics[scale=0.55]{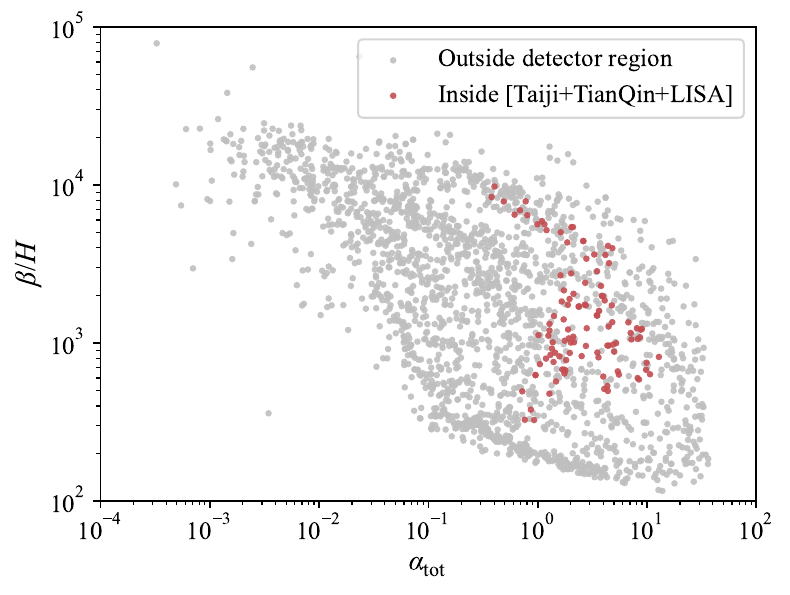}}\qquad
	\subfigure[MC-DTPoint $\sim 32$~hours]{
	\includegraphics[scale=0.55]{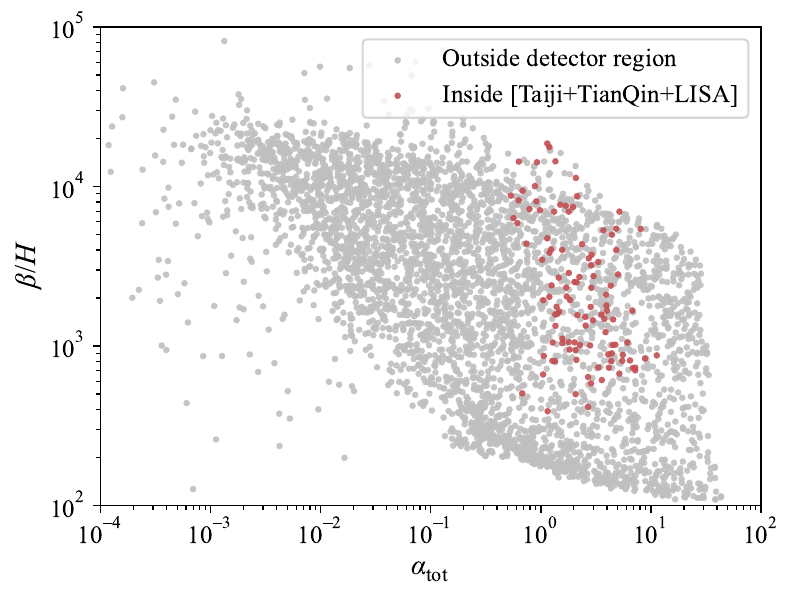}}
    \caption{Comparison of the macroscopic phase transition quantities obtained from the PPO and MC scans, shown in the plane of the phase transition strength $\alpha$ versus the inverse duration parameter $\beta/H$. Panel (a) shows the PPO-GS result, while panel (b) shows the MC-GSPoint result, with all benchmark points color-coded by $T_p/m_{A^\prime}$. Panel (c) shows the PPO-DT result, while panel (d) shows the MC-DTPoint result. In panels (c) and (d), benchmark points falling within the sensitive regions of space-based gravitational wave detectors are marked in red, whereas points outside the detector sensitive regions are shown in light gray.}
    \label{Fig:Macro1}
\end{figure}

We next consider the same symmetry breaking scale window, but now using the detector target reward. The results are summarized in the last two rows of Table~\ref{Tab:NWScan1}. To obtain the same number of potentially detectable benchmark points in the sensitive regions of space-based gravitational wave detectors, the MC scan requires about 4 times as much computational time as the PPO detector target scan.
The corresponding benchmark points are shown in Fig.~\ref{Fig:DT1}.
The figure shows that the MC scan spends a significant amount of computational time sampling detector-irrelevant regions. Thus even with about 4 times the runtime of the PPO detector target scan, MC only reaches the same number of benchmark points within the detector sensitive region.
With the same runtime, MC-Time identifies only 13 points in the detector sensitive regions, as reported in Table~\ref{Tab:NWScan1} and visually illustrated in panel (b) of Fig.~\ref{Fig:GS1}.

We also compare the efficiency of PPO and MC by examining their distributions in terms of the \emph{macroscopic} physical quantities,
namely the phase transition strength $\alpha$ and the inverse duration $\beta/H$
rather than the original microscopic model parameters $\{v_x, g_x, \lambda_x\}$.
The comparisons of PPO-GS with MC-GSPoint and PPO-DT with MC-DTPoint
are shown in Fig.~\ref{Fig:Macro1}.

As observed from Fig.~\ref{Fig:Macro1}, PPO-GS and MC-GSPoint exhibit similar distributions of viable low-temperature benchmark points, shown in purple. This is expected, since MC-GSPoint is designed to collect the same number of viable low-temperature points as PPO-GS. However, outside this target region, the MC scan accumulates a much denser distribution of less relevant points, many of which are physically inconsistent with $T_p/m_{A^\prime}\gtrsim0.5$.
These points are mainly located from the intermediate region toward the left side of the plot.
By contrast, PPO-GS leaves only a sparse footprint in the corresponding region, indicating that it spends substantially less computational effort on unwanted points and is therefore more efficient than the MC scan.

A similar conclusion can be drawn from the comparison between PPO-DT and MC-DTPoint. By construction, the two scans collect the same number of benchmark points within the detector sensitive region, marked in red. Although the full distributions do not exhibit a simple universal pattern, the plots clearly show that the MC scan spends a large fraction of its runtime sampling detector-irrelevant points, leading to a dense distribution outside the targeted region. In contrast, the PPO detector target scan concentrates much more efficiently on the detector sensitive region, providing an effective strategy for locating potentially observable benchmark points.

\subsubsection{Phase transitions at $v_x\in [1,10]~{\rm GeV}$}

We next present the results for $U(1)_x$ phase transitions in the range $v_x\in[1,10]~{\rm GeV}$.
The main results obtained from PPO scans with 80 episodes and 32 steps per episode,
together with the corresponding MC comparison scans, are summarized in Table~\ref{Tab:NWscan10}.
For this parameter region, we perform four PPO scans: the general scan and the detector target scan, as in the previous $v_x$ range, together with their policy transfer variants.
This allows us to test a particularly robust application of reinforcement learning, namely \textbf{PPO policy transfer}, which we denote by ``Tran''. Specifically, we take the PPO-GS and PPO-DT agents trained in the previous section for $v_x\in[0.1,1]~{\rm GeV}$ and apply them directly to the new scale window $v_x\in[1,10]~{\rm GeV}$. We find that the transferred agents perform remarkably better than PPO agents trained from scratch in the new parameter window.

We again compare the PPO results with two MC benchmarks: an MC scan performed with approximately the same runtime, denoted by MC-Time, and an MC scan that finds the same number of viable low-temperature points satisfying $T_p/m_{A^\prime}\lesssim0.2$, denoted by MC-GSPoint. For approximately the same scanning time, PPO-GS finds about 3.12 times as many viable low-temperature phase transition benchmark points as the MC scan. Conversely, to obtain the same number of viable low-temperature phase transition benchmark points as PPO-GS, the MC scan requires about $3.08$ times the PPO-GS computational time. We do not perform a separate MC-DTPoint scan in this $v_x$ range, since MC-GSPoint already identifies a number of detector sensitive benchmark points comparable to that found by PPO-DT.

\begin{figure}
	\centering
	\subfigure[PPO-GS $\sim 8$~hours]{
	\includegraphics[scale=0.46]{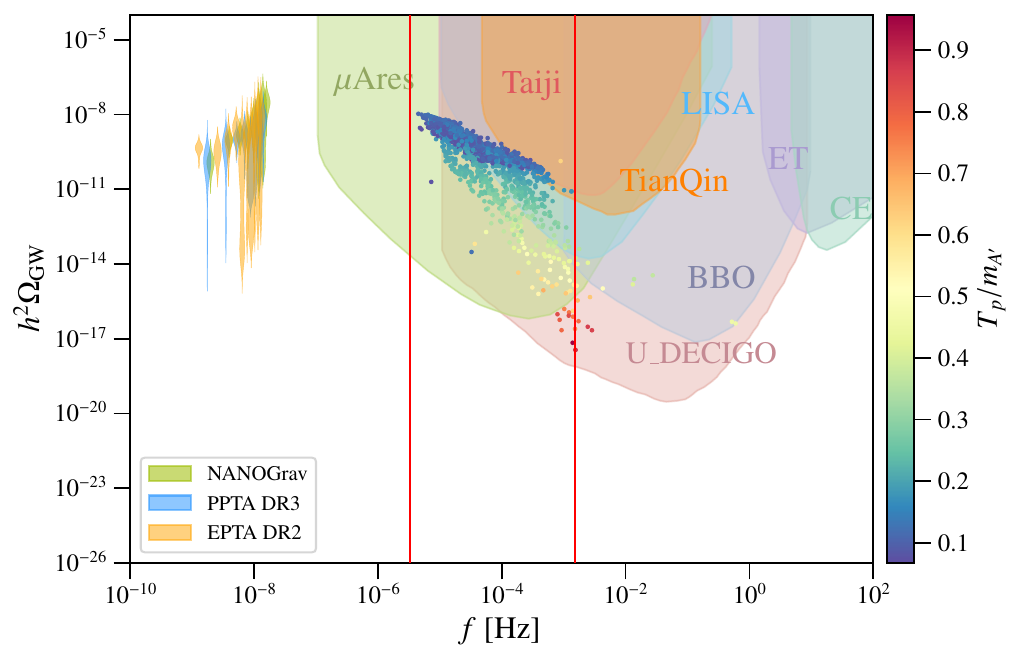}}
	\subfigure[PPO-GS-Tran $\sim 8$~hours]{
	\includegraphics[scale=0.46]{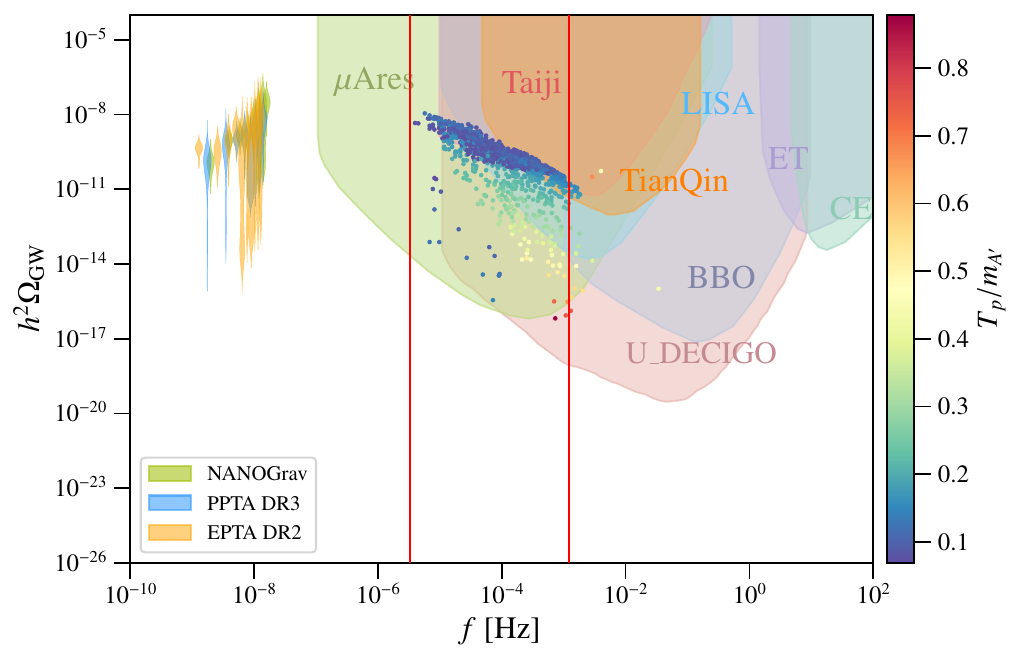}}\\
	\subfigure[MC-Time $\sim 8$~hours]{
	\includegraphics[scale=0.46]{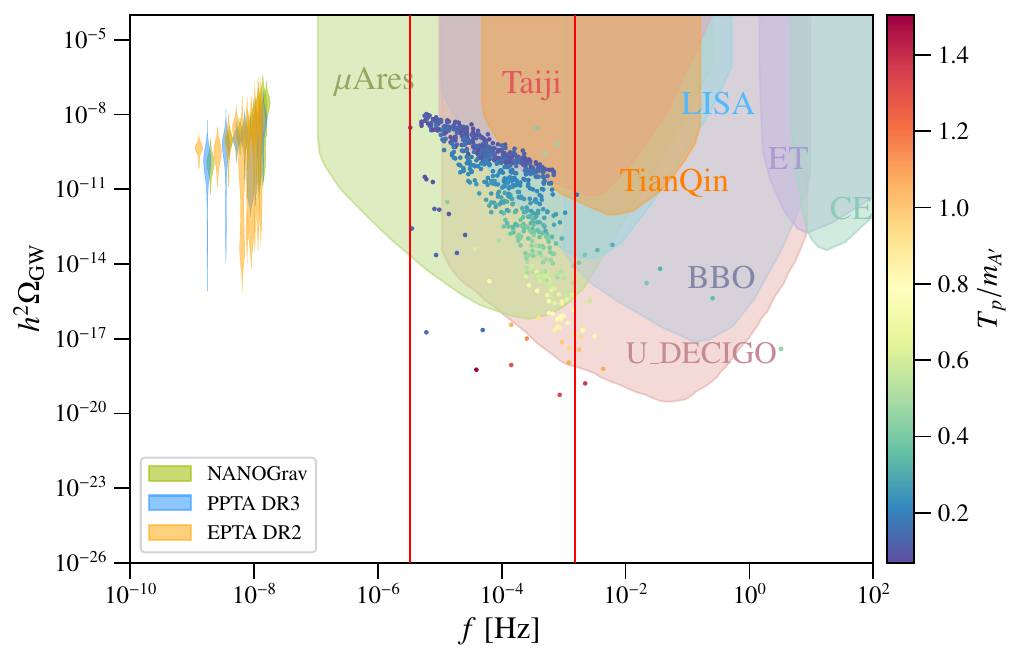}}
	\subfigure[MC-GSPoint $\sim 26$~hours]{
	\includegraphics[scale=0.46]{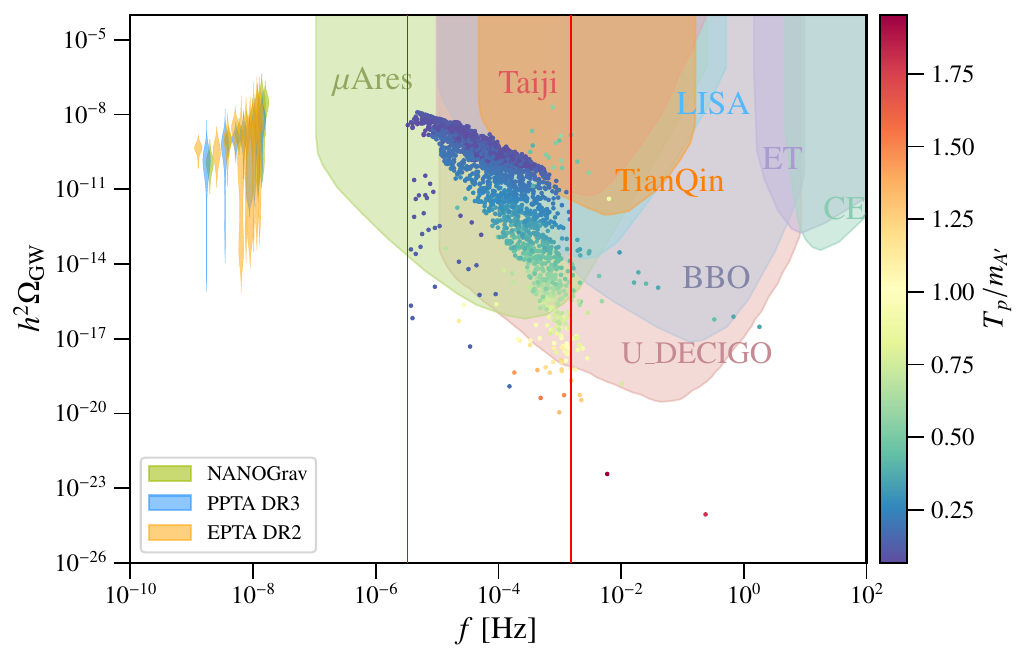}}\\
	\subfigure[PPO-DT $\sim 8$~hours]{
	\includegraphics[scale=0.46]{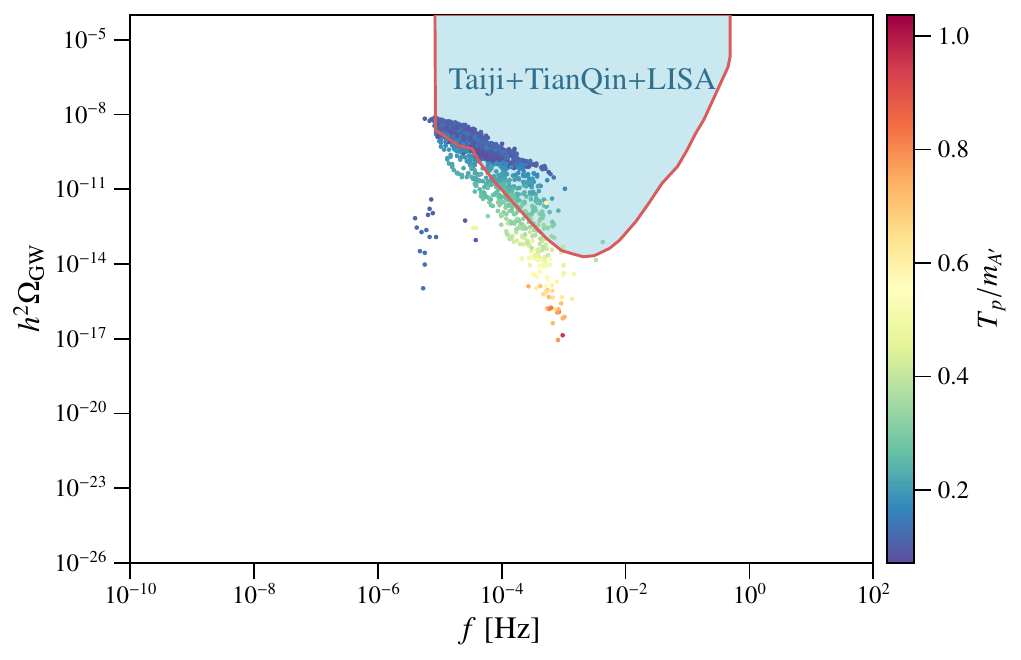}}
	\subfigure[PPO-DT-Tran $\sim 8$~hours]{
	\includegraphics[scale=0.46]{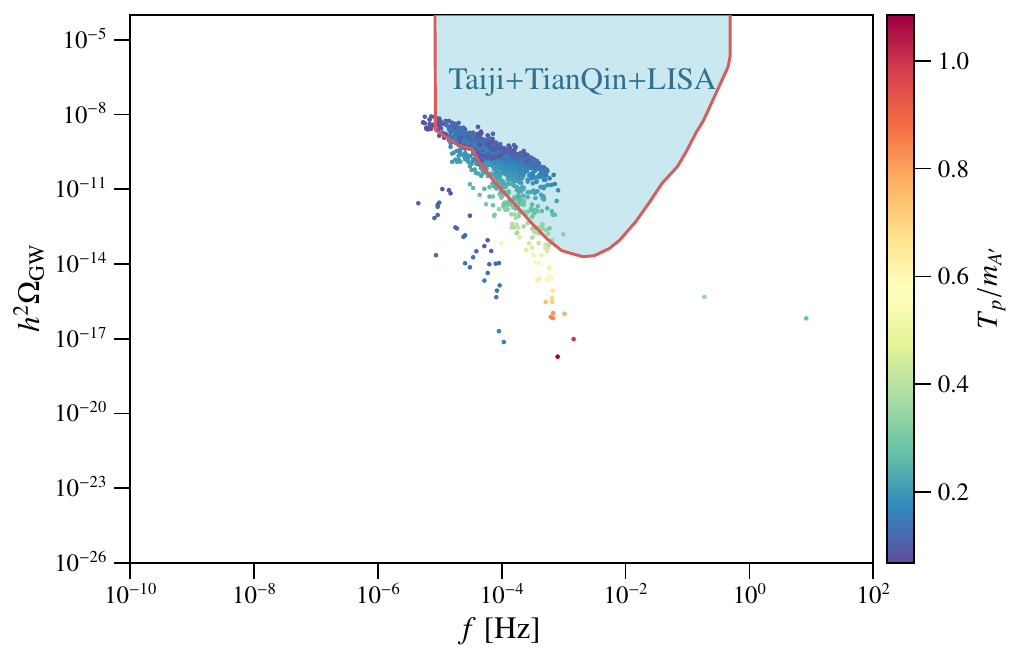}}
\caption{Comparison of PPO and MC scans for the minimal dark $U(1)_x$ model with $v_x\in[1,10]~{\rm GeV}$. All PPO scans are performed for about 8 hours. Panels (a) and (b) show the PPO general scan and the PPO scale transfer scan, respectively. Panels (c) and (d) show the MC scans with the same runtime as PPO-GS and
with the same number of benchmark points satisfying
$T_p/m_{A^\prime}\lesssim 0.2$ as PPO-GS, respectively. The two red vertical lines in panels (a)--(d) are fixed reference lines determined by MC-GSPoint, indicating the frequency range reached by the MC scan after about 26 hours of runtime. Panels (e) and (f) show the PPO detector target scan and the corresponding scale transfer scan, respectively.}
    \label{Fig:GS10}
\end{figure}

The corresponding plots are shown in Fig.~\ref{Fig:GS10}. They demonstrate that PPO is substantially more efficient than the MC scan and also provides a good coverage of the frequency range spanned by the benchmark points. In the $v_x\in[1,10]~{\rm GeV}$ region considered here, a large fraction of benchmark points already fall within the detector sensitive region. As a result, the detector target reward does not lead to a significant improvement over PPO-GS in identifying detector sensitive benchmark points. Nevertheless, given the same search time, PPO-DT still identifies about 3.2 times as many such points as MC-Time.

The \textbf{PPO policy transfer} results demonstrate another powerful aspect of reinforcement learning. PPO-GS-Tran uses the PPO-GS agent trained previously in the $v_x\in[0.1,1]~{\rm GeV}$ region and applies it directly to the new parameter window.
We find that the transferred agent explores an even broader frequency range than the MC scan, which requires about 3.3 times the PPO runtime. This improvement is particularly noticeable in the high-frequency region to the right of the red vertical line.
Similarly, PPO-DT-Tran shows that the agent trained in the $v_x\in[0.1,1]~{\rm GeV}$ region achieves about a $25\%$ improvement in identifying benchmark points within the detector sensitive regions, compared with PPO-DT trained from scratch.

\begin{table}
\center
\begin{tabular}{|c|c|c|c|c|c|}
\hline
$1-10$ & Strategy & Time (s) & Total points & $T_{p}/m_{A^{\prime}}\lesssim0.2$ & Detection\tabularnewline
\hline
\hline
\multirow{4}{*}{PPO} & PPO-GS & 30154 & 1737 & 1311 & 1244\tabularnewline
\cline{2-6}
 & PPO-GS-Tran & 30400 & 1504 & 1292 & 1300\tabularnewline
\cline{2-6}
 & PPO-DT & 30726 & 1717 & 1285 & 1274\tabularnewline
\cline{2-6}
 & PPO-DT-Tran & 30045 & 1835 & 1564 & 1592\tabularnewline
\hline
\multirow{2}{*}{MC} & MC-Time & 30154 & 788 & 412 & 399\tabularnewline
\cline{2-6}
 & MC-GSPoint & 92798 & 2407 & 1311 & 1261\tabularnewline
\hline
\end{tabular}
\caption{Summary of the PPO general scan, PPO detector target scan, PPO policy scale transfer results, and the corresponding MC benchmarks for the minimal dark $U(1)_x$ model with $v_x\in[1,10]~{\rm GeV}$. The PPO category consists of the PPO general scan (PPO-GS) and the PPO detector target scan (PPO-DT), both performed with 80 episodes and 32 steps per episode and requiring about 8 hours of runtime. The PPO policy scale transfer results, denoted by ``Tran'', are also included in this category. For comparison, the MC category contains two benchmark scans: MC-Time scan with approximately the same runtime as PPO; and MC-GSPoint scan that finds the same number of physically viable low-temperature points satisfying $T_p/m_{A^\prime}\lesssim 0.2$. In the table, ``total points'' denotes the total number of benchmark points for which the computation is completed and a phase transition is obtained. Among these points, ``$T_p/m_{A^\prime}\lesssim 0.2$'' counts the number of benchmark points satisfying the low-temperature validity condition, while ``detection'' counts the number of benchmark points that fall within the sensitive regions of space-based gravitational wave detectors.}
\label{Tab:NWscan10}
\end{table}

\begin{figure}
	\centering
	\subfigure[PPO-GS $\sim 8$~hours]{
	\includegraphics[scale=0.54]{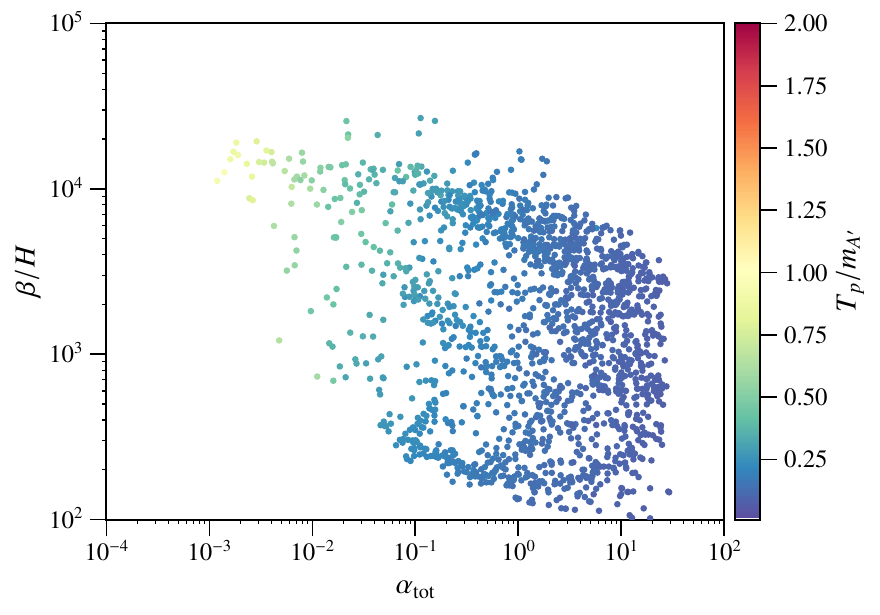}}
	\subfigure[PPO-GS-Tran $\sim 8$~hours]{
	\includegraphics[scale=0.54]{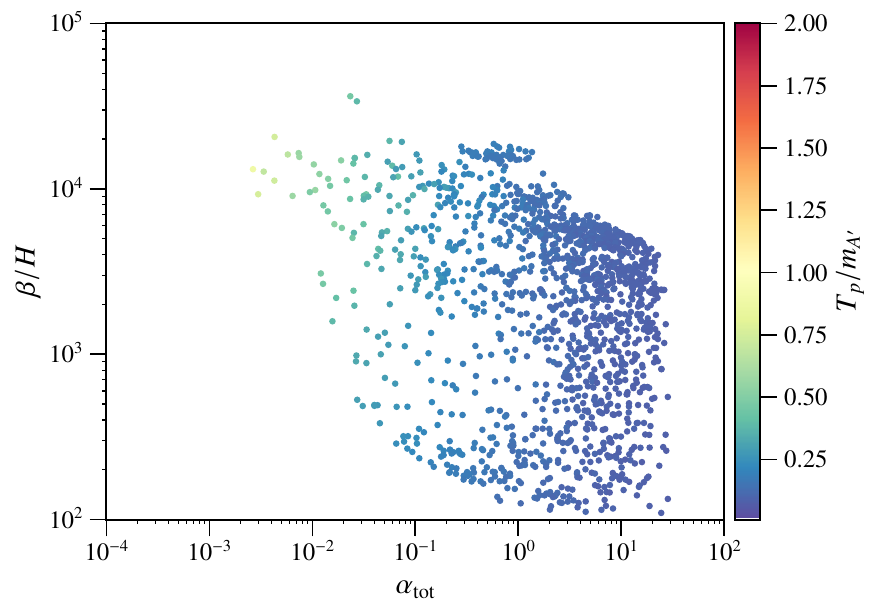}}\\
\hspace{-20pt}
	\subfigure[MC-GSPoint $\sim 26$~hours]{
	\includegraphics[scale=0.54]{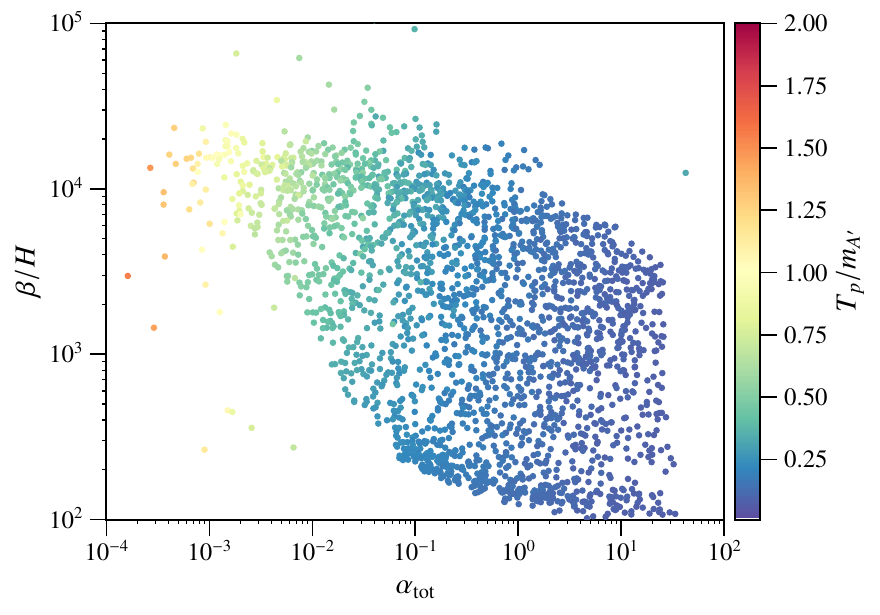}}
	\subfigure[MC-GSPoint $\sim 26$~hours]{
	\includegraphics[scale=0.56]{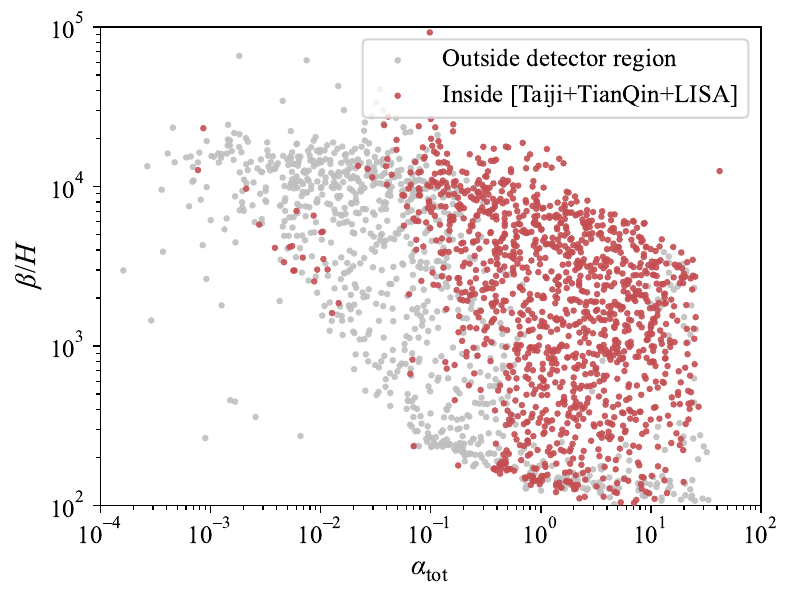}}\\
\hspace{-20pt}
	\subfigure[PPO-DT $\sim 8$~hours]{
	\includegraphics[scale=0.56]{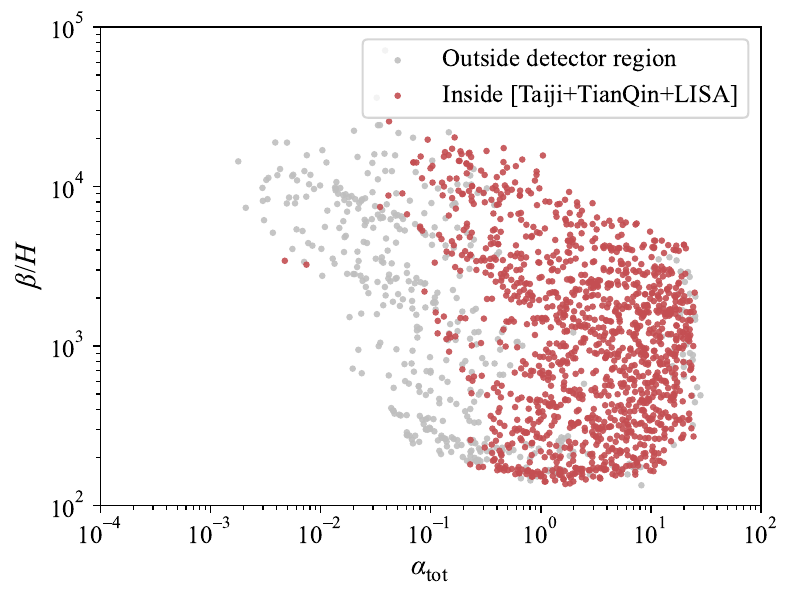}}\qquad
	\subfigure[PPO-DT-Tran $\sim 8$~hours]{
	\includegraphics[scale=0.56]{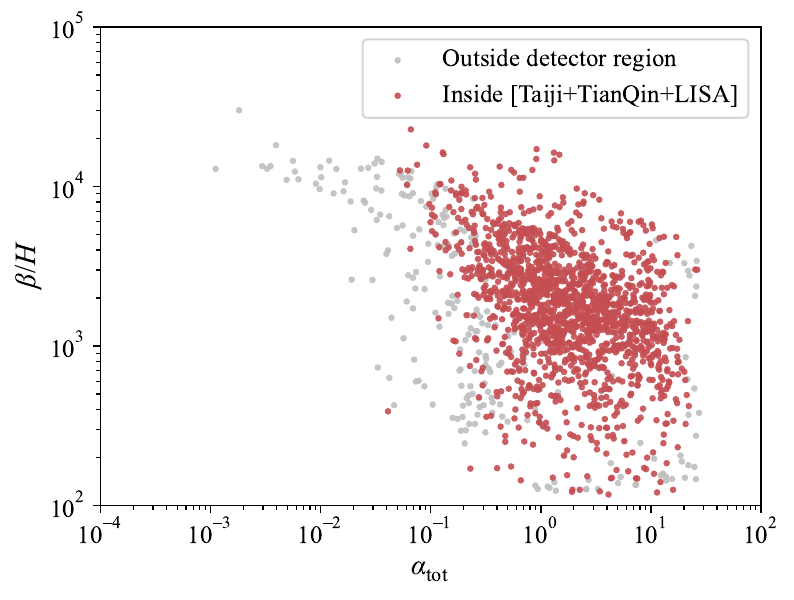}}
    \caption{Comparison of the macroscopic phase transition quantities in the $\alpha$--$\beta/H$ plane obtained from the PPO and MC scans. Panels (a) and (b) show the PPO general scan and the corresponding scale transfer scan, respectively. Panels (c) and (d) show the MC-GSPoint scan results using two different color codings.
    Panels (e) and (f) show the PPO detector target scan and the corresponding scale transfer scan, respectively. In panels (d,e,f), benchmark points within the sensitive regions of space-based gravitational wave detectors are marked in red, while points outside the detector sensitive regions are shown in light gray.}
    \label{Fig:Macro2}
\end{figure}

We further compare the efficiency of PPO and MC by examining their behavior in terms of macroscopic phase transition quantities, namely the transition strength $\alpha$ and the inverse duration $\beta/H$, as shown in Fig.~\ref{Fig:Macro2}. In panels (a) and (b), corresponding to the PPO scans, the low-temperature benchmark points satisfying $T_p/m_{A^\prime}\lesssim 0.2$ are more densely populated, as indicated in purple. By contrast, the MC scan shown in panel (c) yields a more broadly spread distribution across different values of $T_p/m_{A^\prime}$. In panels (e) and (f), a clearer trend is observed: the detector sensitive benchmark points are more strongly clustered, especially for the detector target policy transfer scan. This indicates that the PPO agent learns to improve the search efficiency by exploiting the guidance provided by the reward design.

\subsection{Broad-window scan: $v_x\in[0.1,10]~{\rm GeV}$}

\begin{table}
\center
\begin{tabular}{|c|c|c|c|c|c|}
\hline
$0.1-10$ & Strategy & Time (s) & Total points & $T_{p}/m_{A^{\prime}}\lesssim0.2$ & Detection\tabularnewline
\hline
\hline
\multirow{5}{*}{PPO} & PPO-GS & 57595 & 4028 & 2596 & 778\tabularnewline
\cline{2-6}
 & PPO-HI & 63904 & 4391 & 2727 & 1158\tabularnewline
\cline{2-6}
 & PPO-FB & 55025 & 4380 & 2294 & 387\tabularnewline
\cline{2-6}
 & PPO-DT & 61665  & 4061 & 2137 & 1307 \tabularnewline
\cline{2-6}
 & PPO-DT-Tran & 62311 & 4419 & 3577 & 2412 \tabularnewline
\hline
\multirow{2}{*}{MC} & MC-Time & 58000 & 1256 & 528 & 260\tabularnewline
\cline{2-6}
 & MC-Point & 209045 & 6736 & 2500 & 1225\tabularnewline
\hline
\end{tabular}
\caption{Summary of the five PPO scans and two MC scans for the broad window $v_x\in[0.1,10]~{\rm GeV}$.
The PPO category includes the general scan (GS), history-informed global scan (HI), fixed boundary scan (FB), detector target scan (DT), and the corresponding policy transfer scan (DT-Tran), each performed with 160 episodes and 32 steps per episode, requiring about 16 hours of runtime. For comparison, the MC category includes two benchmark scans: MC-Time, an MC scan performed with approximately the same runtime as PPO, and MC-Point, an MC scan that finds a similar number of low-temperature points satisfying $T_p/m_{A^\prime}\lesssim 0.2$. In the table, ``Total points'' denotes the total number of benchmark points for which the computation is completed and a phase transition is obtained. Among these points, ``$T_p/m_{A'}\lesssim 0.2$'' counts the number of benchmark points satisfying the low-temperature validity condition, while ``Detection'' counts the number of benchmark points that fall within the sensitive regions of space-based gravitational-wave detectors.}
\label{Tab:BWScan}
\end{table}

In this subsection, we compare the MC and PPO scans within the broad vev window $v_x\in[0.1,10]~{\rm GeV}$, which spans two orders of magnitude. We consider all four PPO reward designs in this window.
All PPO broad-window scans are performed with \emph{160 episodes and 32 steps per episode}.

\begin{figure}
	\centering
	\subfigure[PPO-GS $\sim 16$~hours]{
	\includegraphics[scale=0.45]{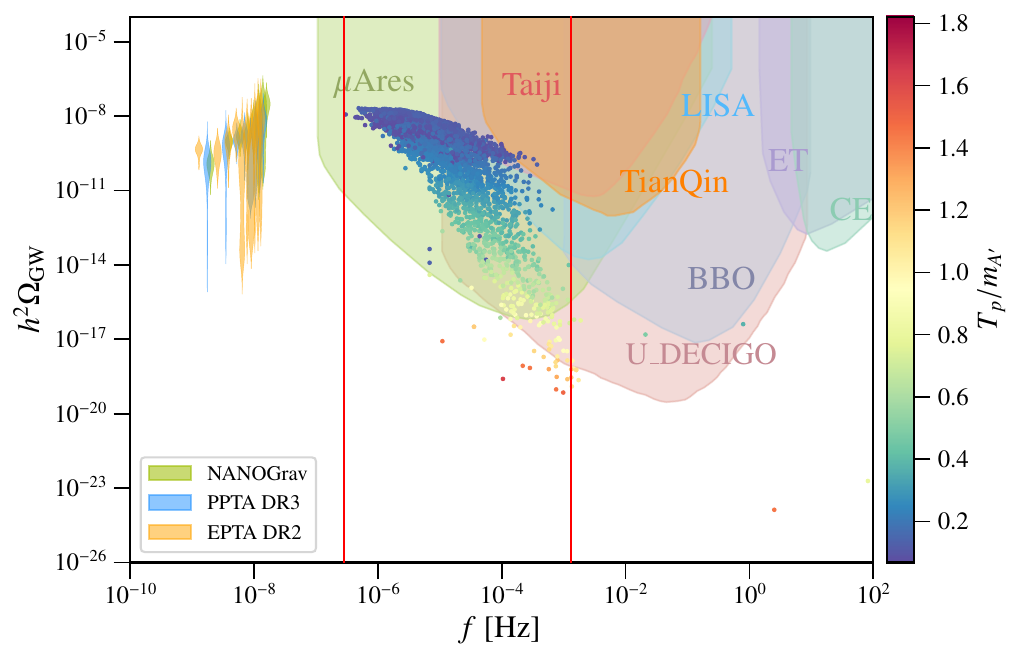}}\quad
	\subfigure[PPO-HI $\sim 16$~hours]{
	\includegraphics[scale=0.45]{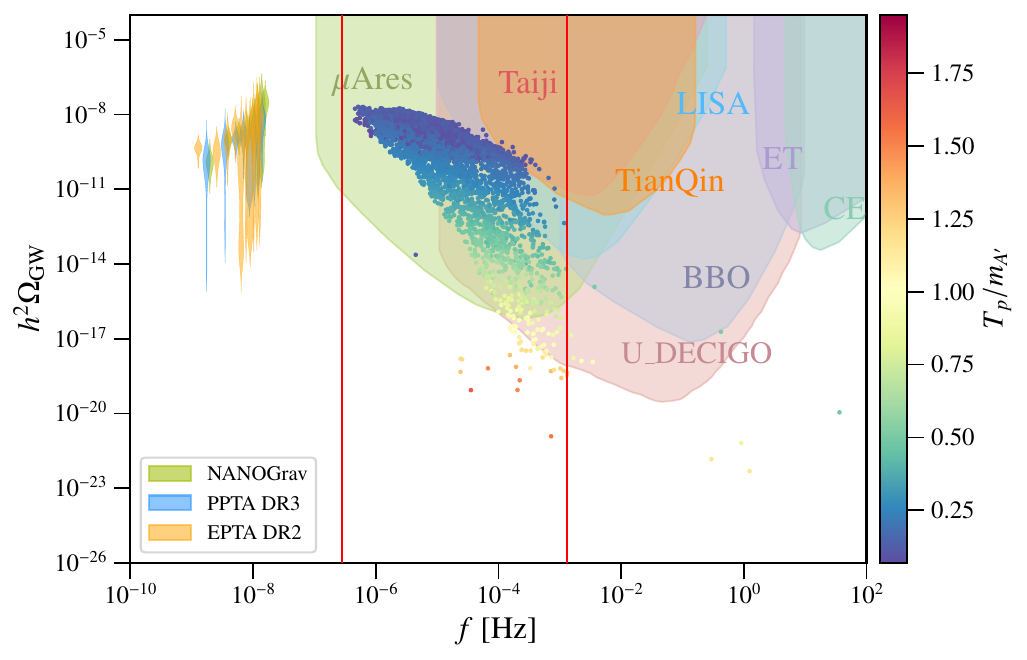}}\\
	\subfigure[PPO-FB $\sim 16$~hours]{
	\includegraphics[scale=0.45]{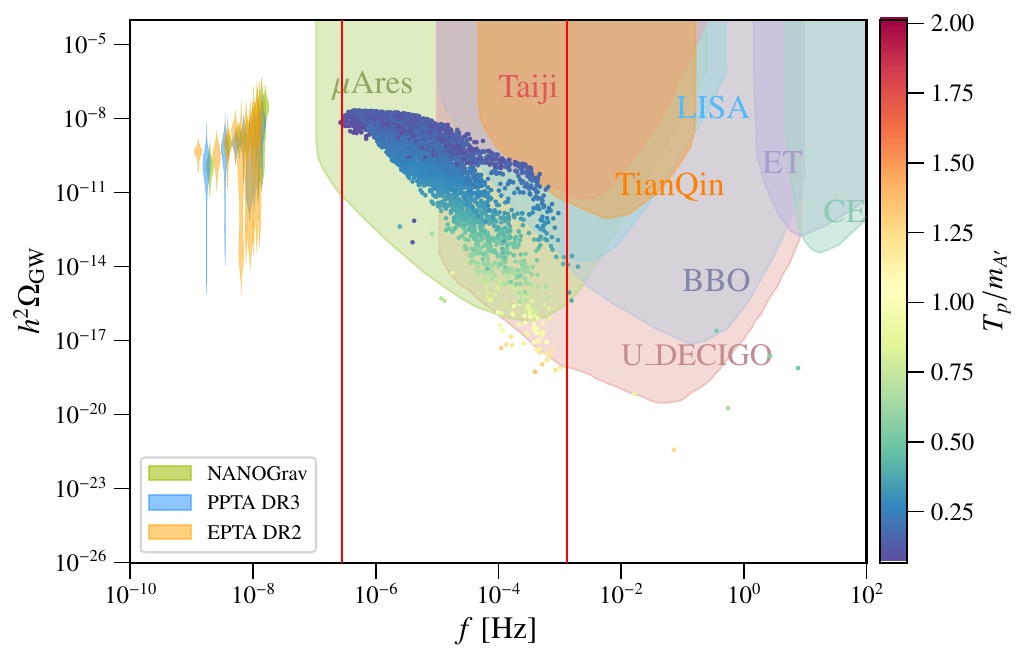}}\quad
	\subfigure[MC-Point $\sim 58$~hours]{
	\includegraphics[scale=0.45]{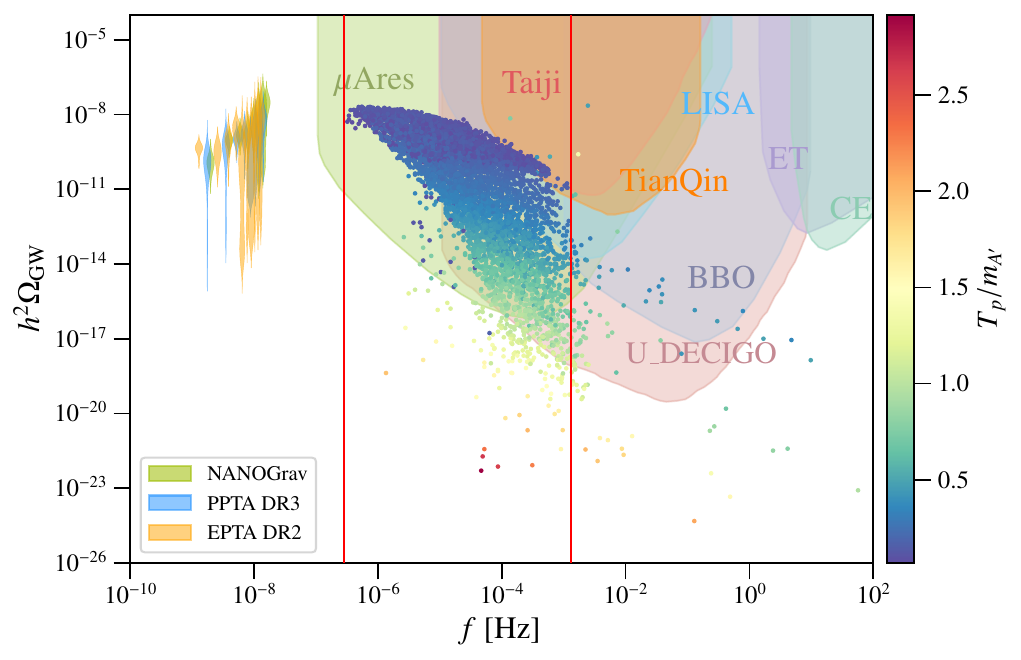}}\\
	\subfigure[PPO-DT $\sim 16$~hours]{
	\includegraphics[scale=0.45]{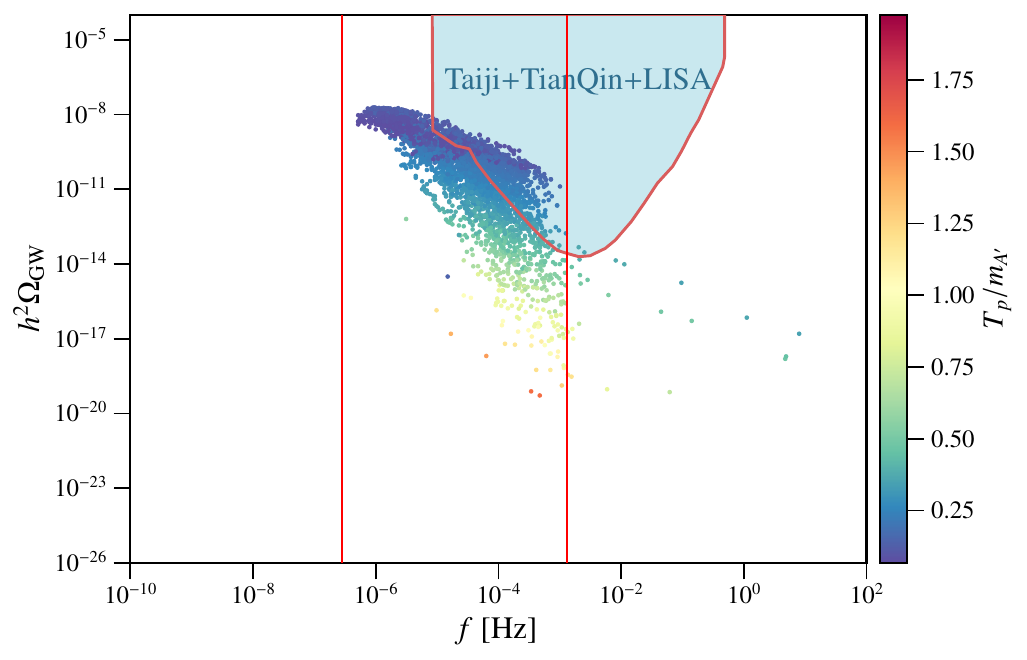}}\quad
	\subfigure[PPO-DT-Tran $\sim 16$~hours]{
	\includegraphics[scale=0.45]{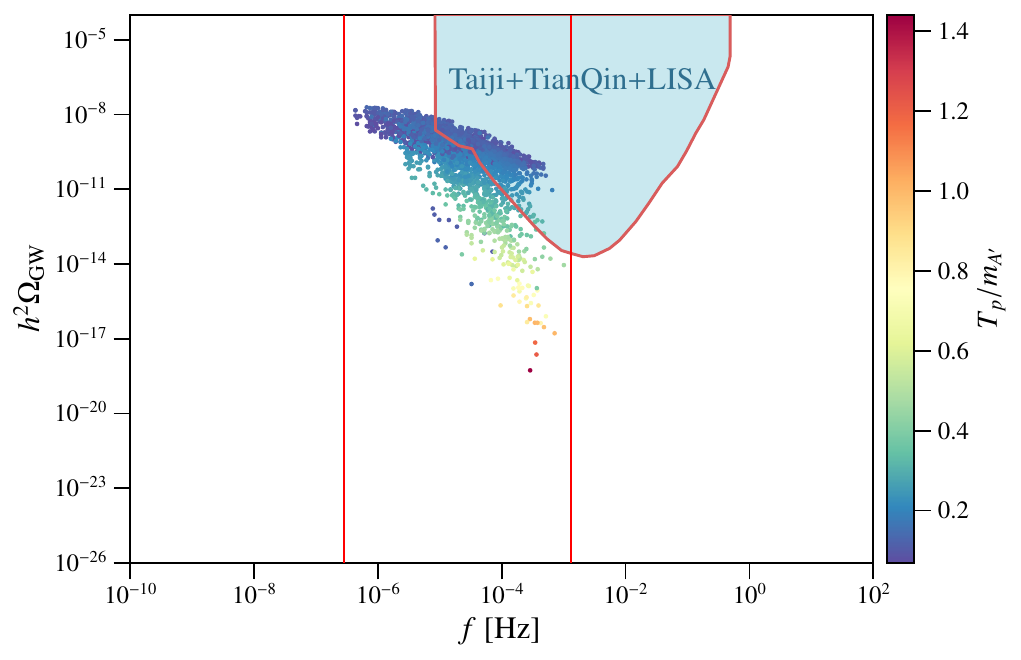}}
\caption{Comparison of PPO and MC scans for the minimal dark $U(1)_x$ model in the broad window $v_x\in[0.1,10]~{\rm GeV}$. All PPO scans are performed for about 16 hours. Panels (a,b,c,e,f) show the PPO results for the general scan, the history-informed global scan, the fixed boundary scan, the detector target scan, and the scale transfer of the detector target policy trained in the narrow window, respectively. Panel (d) shows the MC scan that finds a similar number of benchmark points satisfying $T_p/m_{A^\prime}\lesssim 0.2$ as PPO-GS. The two red vertical lines in all panels are fixed reference lines determined by MC-Point, indicating the frequency range reached by the MC scan after about 58 hours of runtime.}
    \label{Fig:BW}
\end{figure}

The main results are summarized in Table~\ref{Tab:BWScan}. For the broad-window region, we perform five PPO scans: the general scan (GS), the history-informed global scan (HI), the fixed boundary scan (FB), the detector target scan (DT), and the corresponding policy-transfer scan (DT-Tran). Each PPO scan runs for about 16 hours. For comparison, we also perform two MC benchmark scans: an MC scan performed with approximately the same runtime as PPO-GS (MC-Time), and an MC scan that finds a similar number of viable low-temperature points satisfying $T_p/m_{A'}\lesssim0.2$ (MC-Point). The corresponding plots are shown in Fig.~\ref{Fig:BW}.

We find that, for the same scanning time, the PPO scans are generally more efficient than the MC scan. However, for the broader objective of identifying benchmark points with large gravitational wave amplitudes while maintaining broad frequency coverage,
the PPO performance in the broad window is not, as expected, uniformly superior to that of the sufficiently long MC scan.
Among the four PPO scan strategies, the PPO general scan does not provide frequency coverage as broad as the MC-Point scan.
The history-informed global scan performs better than the general scan in this broad-window region,
but it still does not achieve frequency coverage as broad as that of the MC-Point scan.
The fixed boundary scan improves the frequency coverage, but for the same scanning time the inner region of the parameter space is still not fully explored. By contrast, the detector target scan shows the best overall performance. It is highly efficient in locating benchmark points inside the detector sensitive region, while also identifying points with relatively large gravitational wave amplitudes and maintaining broad frequency coverage.

The PPO policy transfer scan, obtained by taking the detector targeted policy trained in the narrow window $v_x\in[0.1,1]~{\rm GeV}$ and applying it directly to the broad window $v_x\in[0.1,10]~{\rm GeV}$, shows excellent performance in identifying benchmark points within the detector sensitive region. It finds nearly twice as many detector sensitive benchmark points as the detector target scan trained from scratch, and also outperforms the MC scan even though the MC scan is run for about $3.6$ times the runtime of the policy transfer scan. However, the frequency coverage of the policy transfer scan does not extend as far toward the right boundary as either the MC scan or the detector target scan trained from scratch.

\begin{figure}
	\centering
	\subfigure[MC-Point $\sim 58$~hours]{
	\includegraphics[scale=0.56]{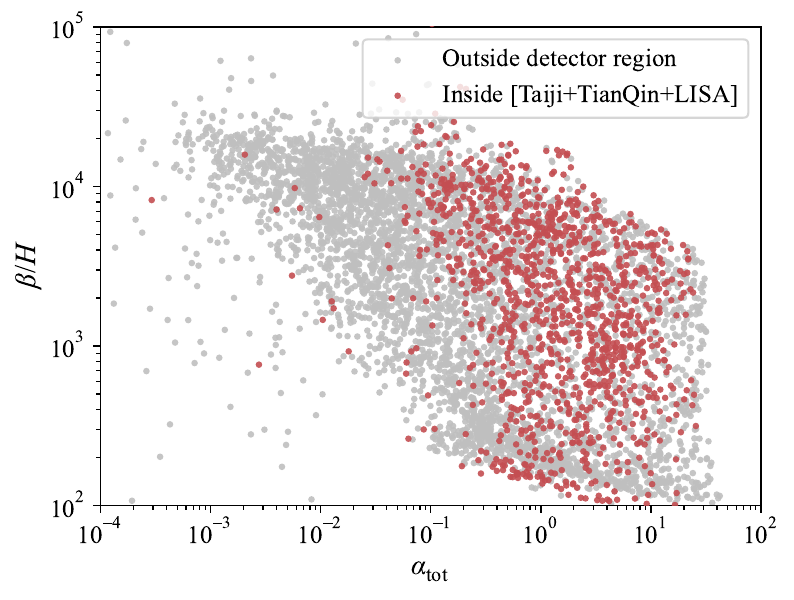}}\\
	\subfigure[PPO-DT $\sim 16$~hours]{
	\includegraphics[scale=0.56]{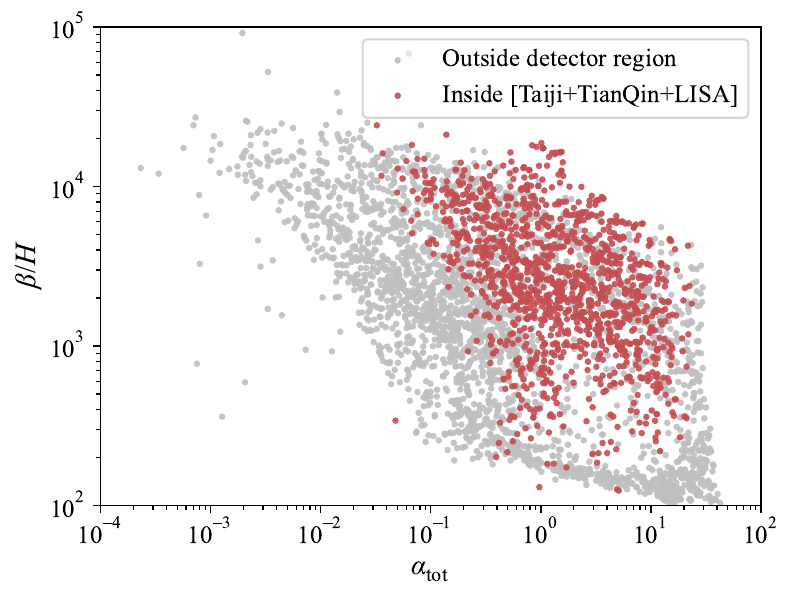}}\qquad
	\subfigure[PPO-DT-Tran $\sim 16$~hours]{
	\includegraphics[scale=0.56]{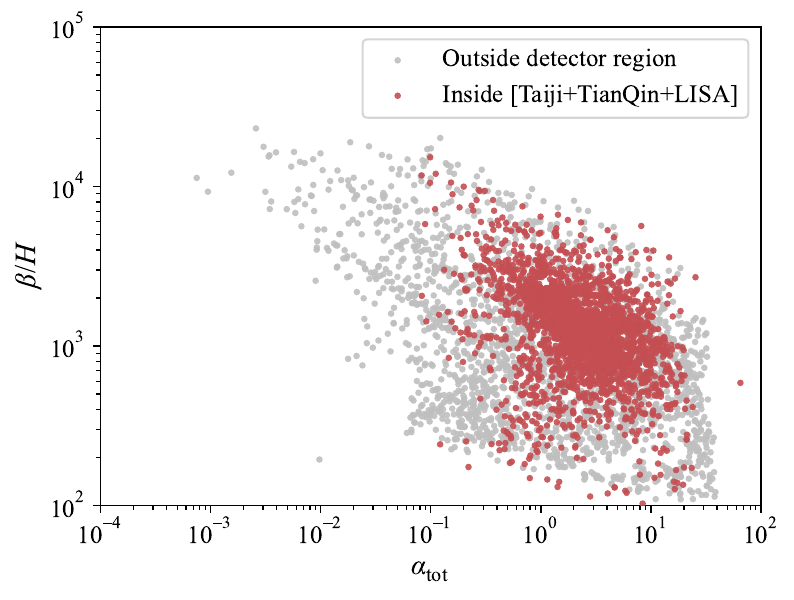}}
\caption{Comparison of the macroscopic phase transition quantities in the $\alpha$--$\beta/H$ plane obtained from the PPO and MC scans.
Panel (a) shows the MC-Point scan result with about 58 hours of runtime, while panels (b) and (c) show the PPO detector target scan and the corresponding scale transfer scan, respectively. Benchmark points within the sensitive regions of space-based gravitational wave detectors are marked in red, whereas points outside the detector sensitive regions are shown in light gray.}
    \label{Fig:Macro3}
\end{figure}

We also compare the efficiency of PPO and MC scans by examining their behavior in terms of the macroscopic phase transition quantities $\alpha$ and $\beta/H$, as shown in Fig.~\ref{Fig:Macro3}. The detector sensitive benchmark points are more efficiently populated in the PPO detector target scan, and become even more concentrated in the corresponding policy transfer scan. This demonstrates that PPO successfully guides the exploration toward the detector sensitive regions of the parameter space.

\subsection{Interpretation of scan performance}

The relative performance of the PPO scan strategies depends sensitively on the scanned range of the $U(1)_x$ vev. We emphasize that the detector target scan used in this work is not designed solely to locate points inside the detector sensitive region. Rather, its reward also contains the amplitude-based reward of the general scan, supplemented by additional reward components that favor benchmark points in the sensitive regions of space-based gravitational wave detectors. Therefore, the detector target scan retains the ability to search for large amplitude gravitational wave signals, while also being explicitly guided toward experimentally relevant detector frequency ranges.

In the narrow-window scans, where $v_x$ varies over one order of magnitude, we find that the \emph{general scan} provides the best performance for the broad objective of identifying large amplitude benchmark points while maintaining wide frequency coverage. This can be understood from the fact that the accessible frequency range is already relatively limited in a narrow window. In this case, the local adaptive reward used in the general scan is sufficient to guide the agent efficiently. Since the reward favors points whose peak amplitudes are large relative to the best amplitude previously found in a nearby frequency region, the agent is encouraged to populate different local frequency regions with strong gravitational wave signals without imposing an additional boundary-seeking bias.

By contrast, the history-informed global scan and the fixed boundary scan are less efficient in the narrow-window case.
The history-informed global scan extends the general scan by keeping track of the lowest and highest frequency boundary points.
However, because the reward design is more complicated and the scan must update the historical frequency boundaries during training,
the available runtime of about $8$~hours is not sufficient for PPO to achieve stable and efficient learning.
As a result, the history-informed global scan does not provide a clear advantage over the general scan in the narrow-window region.
The fixed boundary scan may suffer from a different limitation. If the prescribed frequency and amplitude reference values are not well aligned with the accessible region, the reward may bias the agent toward a preferred subregion rather than improving the overall coverage of the narrow window. On the other hand, the detector target scan performs very well for its intended purpose, efficiently locating benchmark points that fall within the detector sensitive regions.

For the broad-window scan, $v_x\in[0.1,10]~{\rm GeV}$, the situation changes. In this case, the variation of $v_x$ opens a much wider range of possible peak frequencies, making the learning task more challenging. We find that, for the same scanning time, the PPO scans are generally more efficient than the MC scan. However, for the broader objective of identifying benchmark points with large gravitational wave amplitudes while maintaining broad frequency coverage, the PPO performance in the broad window is not uniformly superior to that of a sufficiently long MC scan.

Among the PPO scan strategies, the general scan remains efficient at finding locally strong signals, but it does not provide frequency coverage as broad as the long MC-Point scan. The history-informed global scan performs better than the general scan in this broad-window region, although its frequency coverage is still not as broad as that of the sufficiently long MC scan. With the longer runtime of about $16$~hours, PPO achieves better frequency coverage than in the general scan as the history-informed global scan is designed to encourage global frequency exploration. However, its reward depends strongly on accumulated historical information, which can make the learning process more demanding. The fixed boundary scan improves the frequency coverage by providing a stable and history-independent target, but for the same scanning time the inner region of the parameter space is not fully explored.

By contrast, the detector target scan shows the best overall performance in the broad-window case, since its reward combines the general scan amplitude reward with additional detector-related reward components.
The detector target scan is thus able to identify benchmark points with relatively large gravitational wave amplitudes while also efficiently locating points inside the detector sensitive regions. It therefore performs well not only as a detector-focused scan, but also as a practical scan strategy for maintaining broad frequency coverage in the broad-window parameter space.

We also test a policy transfer strategy by taking the detector target policy trained in the narrow window $v_x\in[0.1,1]~{\rm GeV}$ and applying it directly to the broad window $v_x\in[0.1,10]~{\rm GeV}$. This policy transfer scan shows excellent performance in identifying detector sensitive benchmark points, outperforming both the detector target scan trained from scratch and the sufficiently long MC scan. However, its frequency coverage does not extend as far toward the right boundary as either the long MC scan or the detector target scan trained from scratch. This indicates that policy transfer can substantially improve detector-focused exploration, but may also inherit biases from the narrower training window.

These results show that there is no universally optimal reward prescription. For narrow parameter windows, the local adaptive reward used in the general scan is sufficient and gives the best frequency coverage while efficiently locating large amplitude points. For broad parameter windows, the detector target reward becomes more effective because it combines amplitude-based exploration with explicit guidance toward experimentally relevant detector regions. Policy transfer can further enhance the efficiency of detector-focused searches, although additional mechanisms may be needed to recover the full frequency coverage of the broader parameter space.

\section{Conclusion and outlook}\label{Sec:Con}

The success of AlphaGo naturally raised a broader question: beyond achieving strong performance, can an AI system reveal useful strategies through its decisions? In other words, can we learn not only from what the AI finds, but also from how it explores and prioritizes different possibilities?

The main motivation of the present work is twofold. First, one of the most direct applications of AI to science is to improve the efficiency of existing computational tools. One may argue that conventional studies of gravitational waves from cosmological phase transitions, especially MC scans of the model parameter space, are not prohibitively slow given current computing power and parallelized multi-core computations. Our goal, however, is to address a more general question: when a scientific problem has a clearly defined objective, can AI accelerate the research process? In this sense, our study uses a concrete problem in gravitational wave physics as a modest starting point to demonstrate how reinforcement learning can be incorporated into realistic scientific workflows.

Second, we find that an equally important aspect of this work is to understand how the reinforcement learning agent responds to the reward structure we design. The performance of the PPO scan is not determined solely by the algorithm itself, but also by how the scientific objectives are translated into reward functions. By examining the behavior of the trained agent, we can assess whether the reward captures the intended physics goal and then refine the reward design accordingly.

Returning to the question raised at the beginning, the strategic structure of a game such as Go may be too complex for us to fully understand the long-term reasoning behind an AI-generated sequence of moves. By contrast, in reinforcement learning applications with relatively simple and physically motivated rewards, it is both possible and useful to analyze how the agent interprets the reward system. This interplay between reward design, agent behavior, and scientific objective is one of the most valuable lessons of the present study.

This paper presents three main developments: (1) the construction of a reinforcement learning environment for gravitational wave predictions from cosmological phase transitions, (2) a PPO-based scanning strategy with detector-motivated reward design, and (3) a quantitative comparison with standard MC scanning in terms of both efficiency and the quality of the discovered benchmark points. We demonstrate that PPO scanning, as a form of \emph{goal-directed discovery}, can substantially improve the efficiency of identifying benchmark points that produce large gravitational wave signals in the target frequency band, while maintaining control over the relevant physical constraints. In gravitational wave physics, this approach is well aligned with the practical goals of phenomenological scans, especially when the purpose is to identify representative benchmarks, reveal correlations among macroscopic phase transition parameters, and clarify which microscopic structures preferentially lead to detectable gravitational wave signals.

Beyond the specific model studied here, our framework illustrates a more general paradigm: \emph{scientific exploration as goal-directed, learning-assisted search}, complementary to traditional unbiased sampling. As gravitational wave detectors mature and theoretical models proliferate, such methods will become increasingly valuable for connecting broad beyond-SM landscapes to concrete observational targets.

More broadly, the methodology developed in this paper is applicable across scientific disciplines and research directions. Many modern problems in physics, chemistry, materials science, biology, climate science, and engineering share a common computational structure: a large and often high-dimensional design or hypothesis space, an expensive and sometimes fragile forward model, and a discovery objective that is intrinsically \emph{rare-event} or \emph{goal-directed}. In such settings, learning-based agents can serve as adaptive search engines that allocate computational effort to regions of greatest scientific value, thereby enabling a more direct connection between theoretical possibilities and experimentally actionable targets. Modern computational science methods can thus guide expensive calculations toward the most informative or impactful regions of the problem space, while traditional methods retain their essential role in statistical interpretation, uncertainty quantification, and global characterization.\\

\noindent\textbf{Acknowledgments:}

WZF has benefited from the ``Frontiers in Quantum Physics'' Symposium
organized by the School of Physics and the Center for High Energy Physics at Peking University,
where WZF first conceived the idea of applying reinforcement learning to the present work.
This work is supported in part by the National Natural Science Foundation of China under Grant No. 11935009,
and Tianjin University Self-Innovation Fund Extreme Basic Research Project Grant No. 2025XJ21-0007.

\appendix

\section{PPO Notation and Terminology}
\label{App:PPO}

For clarity, we summarize the PPO notation used in this work.

\medskip

\noindent
\textbf{Step index $t$\,:\ }
The label $t$ denotes the discrete time step within an episode.

\medskip

\noindent
\textbf{State $s_t$\,:\ }
The state $s_t$ denotes the information provided by the numerical environment to the agent at step $t$. In this work, it encodes the relevant model parameters and, depending on the scan strategy, additional information about the gravitational wave signal.

\medskip

\noindent
\textbf{Action $a_t$\,:\ }
The action $a_t$ denotes the move in parameter space proposed by the agent at step $t$. It determines how the model parameters are updated from the current state to the next candidate benchmark point.

\medskip

\noindent
\textbf{Reward $r_t$\,:\ }
The reward $r_t$ is the scalar feedback assigned by the numerical environment after the agent takes action $a_t$.
It is designed to quantify how well the generated parameter point satisfies the search objective.

\medskip

\noindent
\textbf{Episode length $T$\,:\ }
The quantity $T$ denotes the \emph{number} of steps in one episode. In the present application, one episode corresponds to one completed search trajectory in the environment.

\medskip

\noindent
\textbf{Discount factor $\gamma$\,:\ }
The parameter $\gamma$ is the discount factor used in the discounted return. It controls the relative importance of immediate and future rewards.

\medskip

\noindent
\textbf{Policy $\pi_{\theta}(a_t|s_t)$\,:\ }
The policy $\pi_{\theta}(a_t|s_t)$ denotes the probability of selecting action $a_t$ when the agent is in state $s_t$. The subscript $\theta$ denotes the trainable parameters of the policy network.

\medskip

\noindent
\textbf{Policy parameters $\theta$\,:\ }
The symbol $\theta$ collectively denotes the trainable weights and biases of the policy network. In practice, $\theta$ represents a high-dimensional set of parameters.

\medskip

\noindent
\textbf{Old policy $\pi_{\theta_{\rm old}}(a_t|s_t)$\,:\ }
The old policy $\pi_{\theta_{\rm old}}$ denotes the fixed policy used to generate the sampled trajectories stored in the rollout buffer. It is used as the reference policy during the PPO update.

\medskip

\noindent
\textbf{Value function $V^{\pi_{\theta}}_{\phi}(s_t)$\,:\ }
The value function $V^{\pi{\theta}}{\phi}(s_t)$ denotes the value network estimate of the expected return from state $s_t$ under the policy $\pi_{\theta}$. The subscript $\phi$ denotes the trainable parameters of the value network.

\medskip

\noindent
\textbf{Value network parameters $\phi$\,:\ }
The symbol $\phi$ collectively denotes the trainable weights and biases of the value network.

\medskip

\noindent
\textbf{Estimated discounted return $\hat{G}_t$\,:\ }
The quantity $\hat{G}_t$ denotes the estimated discounted return constructed from the rewards collected along the sampled trajectory.

\medskip

\noindent
\textbf{Advantage estimate $\hat{A}_t$\,:\ }
The quantity $\hat{A}_t$ denotes the estimated advantage function.
It measures whether the action taken at step $t$ leads to a return larger or smaller than the value expected from the current state.

\medskip

\noindent
\textbf{Probability ratio $\rho_t(\theta)$\,:\ }
The probability ratio $\rho_t(\theta)$ compares the probability of selecting the same action $a_t$ in the same state $s_t$ under the current policy and the old policy.
It enters the PPO clipped surrogate objective and controls the size of the policy update.

\medskip

\noindent
\textbf{Clipping parameter $\epsilon$\,:\ }
The parameter $\epsilon$ is the PPO clipping parameter. It restricts the probability ratio to a finite interval and prevents excessively large policy updates.

\medskip

\noindent
\textbf{Empirical average $\mathbb{E}_t[\cdots]$\,:\ }
The notation $\mathbb{E}_t[\cdots]$ denotes the empirical average over the sampled time steps stored in the rollout buffer.

\medskip

\noindent
\textbf{Value function loss $L^{\rm VF}(\phi)$\,:\ }
The quantity $L^{\rm VF}(\phi)$ denotes the value-function loss minimized to train the value network. The superscript ``VF'' stands for value function.

\medskip

\noindent
\textbf{Clipped surrogate objective $L^{\rm CLIP}(\theta)$\,:\ }
The quantity $L^{\rm CLIP}(\theta)$ denotes the PPO clipped surrogate objective used to update the policy network. Although the symbol $L$ is used, this objective is maximized with respect to the policy parameters $\theta$.

\medskip

\noindent
\textbf{Rollout buffer\,:\ }
The rollout buffer is the temporary storage of sampled trajectories collected before each PPO update. It stores the states, actions, rewards, value estimates, old-policy probabilities, and terminal-state information needed to compute the discounted returns, advantages, and PPO probability ratios.

\medskip

\noindent
\textbf{Environment\,:\ }
The environment denotes the scan procedure that evaluates the benchmark point generated by the action $a_t$. It computes the relevant phase transition and gravitational wave quantities, assigns the reward $r_t$, and constructs the next state $s_{t+1}$.


\end{document}